\documentclass[10pt,twocolumn,letterpaper]{article}

\usepackage{cvpr}              

\usepackage[dvipsnames]{xcolor}

\definecolor{cvprblue}{rgb}{0.21,0.49,0.74}
\usepackage[pagebackref,breaklinks,colorlinks,citecolor=cvprblue]{hyperref}

\usepackage{multirow}
\usepackage{colortbl}
\usepackage{xcolor}
\definecolor{c1}{HTML}{ffcc99}
\definecolor{c2}{HTML}{fff8ae}

\usepackage{graphicx}

\usepackage{amsmath}    
\usepackage{bm}         
\usepackage{mathtools}  

\usepackage{graphicx}   

\usepackage{hyperref}   

\usepackage{physics}    

\usepackage{xcolor}     
\usepackage{booktabs}   
\usepackage{enumitem}   

\usepackage[ruled]{algorithm2e} 

\SetAlFnt{\small}
\SetAlCapFnt{\small}
\SetAlCapNameFnt{\small}
\SetAlCapHSkip{0pt}

\def\paperID{367} 
\def\confName{3DV\xspace}
\def\confYear{2026\xspace}

\title{TextureSplat: Per-Primitive Texture Mapping for Reflective Gaussian Splatting}

\author{%
  Mae Younes \quad \quad Adnane Boukhayma\\
  Inria France
}

\begin{document}
\twocolumn[{
\renewcommand\twocolumn[1][]{#1}
\vspace{-4em}
\maketitle
\vspace{-2.5em}
\includegraphics[width=1.0\linewidth]{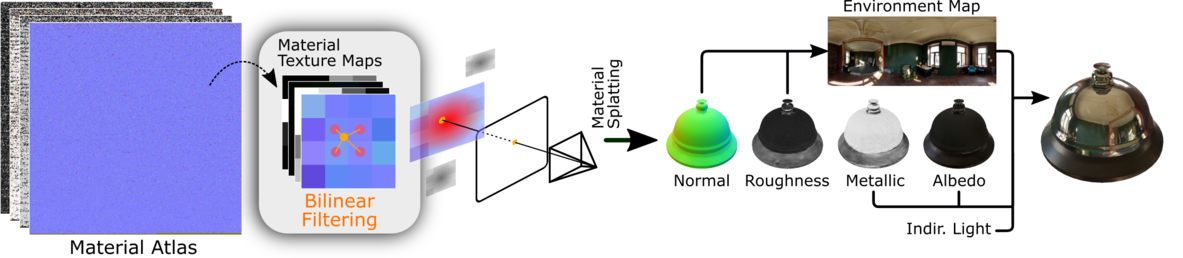}
\vspace{-2em}
\captionof{figure}{{\it Method Overview:} We introduce planar primitive material textures — as opposed to single attributes — within physically based Gaussian Splatting rendering optimization. The increased representation power from spatially varying normal and material in object space enables fidelity reconstruction of high frequency specular in highly reflective scenes. Our hardware-accelerated implementation using texture atlases improves rendering efficiency at test time.
\vspace{1em}}
\label{fig:teaser}
}]
\begin{abstract}
Gaussian Splatting have demonstrated remarkable novel view synthesis performance at high rendering frame rates. Optimization-based inverse rendering within complex capture scenarios remains however a challenging problem.  A particular case is modelling complex surface light interactions for highly reflective scenes, which results in intricate high frequency specular radiance components.
We hypothesize that such challenging settings can benefit from increased representation power. We hence propose a method that tackles this issue through a geometrically and physically grounded Gaussian Splatting borne radiance field, where normals and material properties are spatially variable in the primitive's local space. Using per-primitive texture maps for this purpose, we also propose to harness the GPU hardware to accelerate rendering at test time via unified material texture atlas. Code will be available at{
\textcolor{magenta}
{\texttt{\href{https://github.com/maeyounes/TextureSplat}{TextureSplat}.}}}
\end{abstract}    
\section{Introduction}

3D reconstruction and inverse rendering from multi-view images are pivotal problems receiving constant interest and investigation from computer vision, graphics and machine learning research communities, with a myriad of direct applications in key industrial domains requiring high-quality 3D modeling and visualization.

Neural Radiance Fields (NeRF)~\cite{nerf} revolutionized the field by representing scenes as continuous neural implicit functions optimized through differentiable volume rendering. Building on this foundation, 3D Gaussian Splatting (3DGS)~\cite{3dgs} reintroduced point-based graphics by replacing neural networks with explicit 3D Gaussian primitives, achieving both real-time rendering and state-of-the-art quality. These primitives are rendered through volume resampling~\cite{ewa}, with their parameters optimized via gradient descent-based differentiable rendering. More recently, 2D Gaussian Splatting (2DGS)~\cite{2DGS} improved multi-view consistency by using planar Gaussian primitives that better align with surfaces.

Despite these advances, accurately representing highly reflective surfaces remains challenging. Reflective objects exhibit complex view-dependent effects that depend on surface normals, material properties, and environmental lighting. Recent methods such as  RefGaussian~\cite{RefGaussian} and 3DGS-DR~\cite{3DGS-DR} attempt to model these effects using per-Gaussian material properties and physically-based rendering approaches. However, they often struggle with high-frequency specular highlights and sharp reflections. We hypothesize that this is due in part to the inherent resolution limitations of using a single attribute value per primitive.

In this paper, we ask the question: \textit{How can we enhance the representation power of Gaussian splatting for reflective scenes while maintaining computational efficiency and leveraging hardware acceleration?} Our key insight is that the planar nature of 2D Gaussian primitives naturally defines a parameterization that can be exploited for texture mapping, enabling us to store spatially varying material properties per primitive.

Inspired by the distinction between Gouraud shading~\cite{gouraud} (constant per-vertex attributes) and Phong shading~\cite{phong} (interpolated attributes) in traditional computer graphics, we introduce per-primitive texture maps for material properties in 2D Gaussian splatting. This approach effectively decouples the geometric representation (Gaussian primitives) from the appearance representation (material textures), allowing us to model high-frequency material variations without increasing the number of primitives.

Our method leverages the closed-form ray-splat intersection of 2DGS to accurately map screen-space pixels to local texture coordinates, enabling proper texture filtering. Crucially, we transform tangential normal maps to world space using the primitive's rotation matrix, analogous to normal mapping in traditional rendering. This enables detailed normal variations across each primitive's surface, significantly enhancing the rendering of specular highlights and reflections.

For efficient rendering after optimization, we pack the primitive textures into atlases that leverage GPU hardware-accelerated texture filtering operations. Our approach is fully compatible with deferred shading pipelines, allowing us to incorporate physically-based rendering models for accurate light interactions.

Through experiments on standard benchmarks for reflective scene reconstruction, we demonstrate that our method outperforms state-of-the-art approaches in terms of both quantitative metrics and visual quality. Our method achieves more accurate reflections and sharper specular highlights while maintaining real-time rendering performance. The benefits extend beyond reflective scenes, as our approach improves rendering quality for standard scenes as well.

Our contributions include:\\
\noindent$\bullet$ A per-primitive texture mapping approach for 2D Gaussian splatting that enhances representation power while maintaining computational efficiency.\\
\noindent$\bullet$ Leveraging a normal mapping technique in the context of Gaussian Splatting that significantly improves the quality of specular reflections.\\
\noindent$\bullet$ A hardware-accelerated implementation using texture atlases that enhances real-time rendering at test time.\\
\noindent$\bullet$ State-of-the-art results on benchmarks for reflective scene reconstruction, demonstrating significant improvements in rendering quality and accuracy.
\section{Related Work}

\subsection{Radiance Fields for 3D Scene Representations}
Neural Radiance Fields \cite{nerf} (NeRFs) have been dominating the 3D shape and appearance modelling recently, based on the astounding success of implicit representations combined with differentiable volume rendering \cite{levoy1988volume,volrend}. They represent scenes using view-dependent radiance and density fields parameterized by MLPs. When density is modeled as a function of a signed distance field, NeRF variants enable more accurate geometry reconstruction 
\cite{neus,volsdf,neuralangelo,SparseCraft,GeoTransfer,voxurf}.
However, multi-scale volume rendering demands frequent MLP evaluations, limiting real-time performance. Grid-based methods \cite{ingp,dvgo,kplanes,plenoxels,tensorrf,sun2024sparse} alleviate this but often struggle with large, unbounded scenes even with level-of-detail grids \cite{nsvf}. Implicit reconstruction has been made more robust to noise and sparse observations, whether from images or point clouds, through the use of generalizable data priors (\eg \cite{pixelnerf,mvsnerf,geonerf,genlf,fssdf,nksr,mixing,robust,convocc}) and a variety of regularization strategies (\eg \cite{regnerf,freenerf,dietnerf,dsnerf,rgbdnerf,dro,nap,sparseocc,robust,digs,igr}).
Gaussian splatting (3DGS) \cite{3dgs} emerged lately as a strong alternative to NeRFs, offering state-of-the-art novel view synthesis and real time rendering frame rates. It extends the elliptical weighted average (EWA) volume resampling framework \cite{ewa,zwicker2001ewa} to inverse rendering, modelling scenes with explicit Gaussian kernel primitives, that can be sorted and rasterized efficiently. Recent extensions of Gaussian splatting include building generalizable models \cite{pixsplat,mvsgaussian,SparSplat,HiSplat}, bundle-adjustment–based formulations \cite{bad,colfree,Sparfels,noposplat}, using higher-dimensional primitives \cite{ndim}, spatiotemporal models \cite{4d}, in addition to several methods to improve density control \cite{3dgsmcmc,steepgs,pixelgs}, 
anti-alising \cite{mipsplat,anal,AA-2DGS}, model compactness \cite{compact,contextgs} and training speed \cite{newton,grendel,speedysplat}.
The 2DGS representation \cite{2DGS} leverages planar 2D primitives instead of volumetric ones (3DGS), and performs precise 2D kernel evaluation in object space as opposed to approximative ones in screen space (3DGS), thus leading to superior geometric modelling and multi-view consistency.    

\subsection{Specular Reflection Modeling for Reflective Scenes}
Both vanilla NeRFs and 3DGS assume low-frequency view dependency. Hence, they can struggle with highly reflective scenes. One strategy to improve in this department is using shading functions that are reflection direction aware \cite{ref_nerf,Ref-NeuS,jiang2024rethinking,nerv,nero}. For instance, Ref-NeRF \cite{ref_nerf} extends NeRFs with a new parameterization for view-dependent radiance and incorporates normal vector regularization. The shading function can be more physically grounded, and this enables additional application such as relighting and material editing. In this regard, other NeRF and GS based methods proposed to model light interaction using the explicit rendering equation with BRDF functions. 
ENVIDR \cite{envidr} Uses environment maps to capture spatially-varying reflections in neural rendering. The next wave of work tackled shading for Gaussian Splatting (\eg \cite{3iGS, GS-ROR, GS-IR, Gshader, 3DGS-DR, R3DG, RefGaussian}). 
GShader \cite{Gshader} applies a simplified shading function on each fragment.  
3DGS-DR \cite{3DGS-DR} introduces deferred shading at pixel level, and  stabilizes the optimization by smoothing out
normal gradients. 

Building on the 2DGS representation, our baseline Ref-Gaussian \cite{RefGaussian} decomposes the scene into geometry, material and lighting through the split-sum approximation of the rendering equation while incorporating an indirect light attribute, enabling inter-reflections while being fast to render. It achieves the state-of-the-art performance on the standard reflective scene novel view synthesis benchmarks.
We propose to enhance its physical material and normal representations through the use of per-primitive textures.

\subsection{Texture Attributes in Gaussian Splatting}
Several works \cite{gstex,BBSplat,texturedgaussiansenhanced3d,SuperGaussians,GaussianBillboards,HDGS} recently introduced the texture attribute representation for Gaussian Splatting as well. They use it to model a view independent component of the color. Differently, we explore this representation for rendering material properties and normals within 2DGS enabled physically based rendering to reconstruct challenging highly reflective scenes. We also propose leveraging texture atlases to enable hardware acceleration at test-time rendering. Closest to our context,  
concurrent work \cite{GTAvatar} manages to encode spatially varying primitive material attributes in a single compact texture map thanks to their pre-fitted template mesh.  
\section{Method}
In this section, we present our approach to enhance 2D Gaussian Splatting (2DGS)~\cite{2DGS} for representing highly reflective 3D scenes. We first provide background on 2DGS, then introduce our per-primitive texture mapping method that enables high-frequency detail on flat Gaussian primitives. Finally, we describe our hardware-accelerated implementation using texture atlases and the physically-based rendering model we use for reflective scenes.
\subsection{Background: 2D Gaussian Splatting}

We build upon 2D Gaussian Splatting (2DGS)~\cite{2DGS}, which represents a scene using oriented planar disks offering improved multi-view consistency compared to 3D Gaussian Splatting~\cite{3dgs}. Each primitive $k$ is characterized by its position $\boldsymbol{p}_k$, tangential vectors $\boldsymbol{t}_{u_k}$ and $\boldsymbol{t}_{v_k}$, and scaling factors $(s_{u_k}, s_{v_k})$. The primitive's normal is defined by the cross product $\boldsymbol{n}_k = \boldsymbol{t}_{u_k} \times \boldsymbol{t}_{v_k}$.

The 2D Gaussian is defined in a local tangent plane in world space with coordinates $(u,v)$, parameterized as: 
\begin{align}
P_k(u,v) = \boldsymbol{p}_k + s_{u_k} \boldsymbol{t}_{u_k} u + s_{v_k} \boldsymbol{t}_{v_k} v.
\end{align}

For a point $(u,v)$ in the local coordinate space corresponding to a given pixel $(x,y)$ (\ie ray-splat intersection), the 2D Gaussian value is evaluated as: 
\begin{equation} 
\mathcal{G}_k(u,v) = \exp \left( -\frac{u^2+v^2}{2} \right). 
\end{equation}

Each splat has also a learnable opacity $o_k$. The final pixel color is computed by alpha-blending all primitives that contribute to the pixel in front-to-back order: 
\begin{equation} 
\boldsymbol{A}(x,y) = \sum_{i=1}^N \boldsymbol{a}_i \alpha_i(x,y) \prod_{j=1}^{i-1} (1 - \alpha_j(x,y)) 
\end{equation} 
where $\alpha_i(x,y) = o_i \mathcal{G}_i(u_i(x,y),v_i(x,y))$ is the effective opacity of the $i$-th primitive at pixel $(x,y)$ and $\boldsymbol{a}_i$ is its attribute (e.g., color), typically represented using spherical harmonics for view-dependent effects.

\subsection{Per-Primitive Texture Mapping} \label{sec:texture_splatting}

While existing Gaussian splatting methods assign a single attribute value per primitive, we observe that the flat nature of 2D Gaussians naturally defines a local parameterization that can be leveraged for texture mapping. This enables encoding higher-frequency spatial detail without increasing the number of primitives, analogous to the distinction between Gouraud shading (constant per-primitive attributes) and Phong shading (interpolated attributes) in traditional rendering.

Rather than representing material properties with a single value per primitive, we define them as texture maps in the splat's local coordinate system: 
\begin{equation}
\boldsymbol{a}_k(x,y) = \boldsymbol{\mathcal{T}}_k(u_k(x,y), v_k(x,y)) 
\end{equation} 
where $\boldsymbol{\mathcal{T}}_k$ is the texture map associated with the $k$-th primitive, and $\boldsymbol{a}_k$ represents any reconstructed attribute.

This approach provides several advantages:\\
\noindent\textbf{Decoupling of geometry and appearance}: By separating the geometric representation (Gaussian primitives) from the appearance details (textures), we can represent complex visual features without increasing the number of primitives.\\
\noindent\textbf{Higher fidelity appearance}: Textures can capture high-frequency detail that would otherwise require many more primitives to represent.\\
\noindent\textbf{Normal mapping}: Instead of using a single normal per primitive, we can store detailed normal maps, significantly improving the rendering of specular effects. 

The 2DGS representation is particularly well-suited for texture mapping because the ray-splat intersection already provides exact $(u,v)$ coordinates in the primitive's local space and thus, enables accurate texture filtering. 

\subsection{Texture Mapping Implementation}

To implement our texture mapping approach, we map the local splat coordinates $(u,v)$ to texture coordinates $(s,t)$ that account for the Gaussian kernel's support. Since the Gaussian kernel effectively drops to zero at approximately $S_{\sigma}=3$ standard deviations, we scale the local coordinates to ensure that the effective support of the Gaussian ($[-S_{\sigma},S_{\sigma}] \times [-S_{\sigma},S_{\sigma}]$) maps to the texture space $[0,1] \times [0,1]$:
\begin{equation}
s = \frac{u + S_{\sigma}}{2S_{\sigma}}, \quad t = \frac{v + S_{\sigma}}{2S_{\sigma}}. 
\end{equation}
The attribute value $\boldsymbol{a}_k(x,y)$ for primitive $k$ at pixel $(x,y)$ is then obtained by bilinear filtering from its texture map $\boldsymbol{\mathcal{T}}_k$:
\begin{equation} 
\boldsymbol{a}_k(x,y) = \text{BilinearFilter}(\boldsymbol{\mathcal{T}}_k, s(u_k(x,y)), t(v_k(x,y))). 
\end{equation}

\subsection{Hardware Acceleration via Texture Atlases} 
\label{sec:hardware_acceleration}
To efficiently render optimized scenes at test time with per-primitive textures, we leverage hardware-accelerated texture filtering by packing individual primitive textures into texture atlases. This approach is inspired by methods like Ptex~\cite{Burley2008Ptex} and seamless texture atlases~\cite{purnomo2004seamless}, adapted to the specific needs of Gaussian splatting. We provide details about Texture Atlas construction and texture sampling in the Supplementary Material (Section 1).

\subsection{Physically-Based Deferred Rendering} \label{sec:inverse_rendering}

Following the deferred rendering approach used in 3DGS-DR~\cite{3DGS-DR} and Ref-Gaussian~\cite{RefGaussian}, we first splat material attributes to screen-space buffers, then apply physically-based shading in a separate pass.

\subsubsection{Material Properties}

Each 2D Gaussian is associated with texture maps for the following material properties: Albedo $\boldsymbol{\lambda} \in [0,1]^3$, Metallic $m \in [0,1]$, Roughness $r \in [0,1]$, Tangent normal $\boldsymbol{n}^t \in [0,1]^3$ representing normal perturbations in the tangent space. For memory efficiency, we encode tangent normals using only two components $(n_x^t, n_y^t)$ and reconstruct the third component at runtime using $n_z^t = \sqrt{\max(0, 1 - (n_x^t)^2 - (n_y^t)^2)}$.

\subsubsection{Normal Mapping}

A key factor in our approach is the use of normal mapping instead of a single normal per primitive. The tangent normal map encodes normal perturbations in the primitive's local coordinate system. These are transformed to world space using:
\begin{equation} 
\boldsymbol{n}_k(x,y) = \boldsymbol{R}_k \cdot \boldsymbol{n}^t_k(x,y) 
\end{equation}
where $\boldsymbol{R}_k$ is the primitive's rotation matrix. This enables detailed normal variations across the surface of each primitive, critical for capturing high-frequency specular effects.

\subsubsection{Attribute Splatting}

We splat the material attributes to screen-space buffers using alpha-blending:
\begin{equation} 
\boldsymbol{X}(x,y) = \sum_{i=1}^N \boldsymbol{x}_i(x,y) \alpha_i(x,y) \prod_{j=1}^{i-1} (1 - \alpha_j(x,y)), 
\end{equation}

where $\boldsymbol{X}$ represents the combined screen-space buffers:
\begin{equation}
\boldsymbol{X} = 
\begin{bmatrix} 
\boldsymbol{\Lambda} \\ 
M \\ 
R \\ 
\boldsymbol{N} \\ 
\boldsymbol{L}_{\text{ind}} 
\end{bmatrix}, \quad
\boldsymbol{x}_i = 
\begin{bmatrix} 
\boldsymbol{\lambda}_i(x,y) \\ 
m_i(x,y) \\ 
r_i(x,y) \\ 
\boldsymbol{n}_i(x,y) \\ 
\boldsymbol{l}^{\text{ind}}_i(x,y). 
\end{bmatrix}
\end{equation}

This deferred approach treats alpha-blending as a smoothing filter, stabilizing the optimization of features sampled from textures and producing more cohesive rendering results compared to shading directly on the Gaussians~\cite{RefGaussian, 3DGS-DR}.

\subsubsection{Physically-Based Shading}

With the aggregated material maps, we apply the rendering equation to compute the outgoing radiance $\boldsymbol{L}_o(x,y,\omega_o)$ in the direction $\omega_o$:
\begin{equation} 
\boldsymbol{L}_o(x,y,\omega_o) = \boldsymbol{L}_d(x,y,\omega_o) + \boldsymbol{L}_s(x,y,\omega_o). 
\end{equation}
The diffuse term writes: 
\begin{equation} 
\boldsymbol{L}_d(x,y,\omega_o) = \frac{\boldsymbol{\Lambda}(x,y)}{\pi} (1-M(x,y)) \mathcal{L}_{\text{env}}^{\text{diffuse}}(\boldsymbol{N}(x,y)),
\end{equation}
where $\mathcal{L}_{\text{env}}^{\text{diffuse}}$ is the pre-integrated diffuse environment irradiance. Following the split-sum approximation, we compute the specular component efficiently:
\begin{equation}
\begin{split}
&\boldsymbol{L}_s(x,y,\omega_o) \approx \text{BRDF}_{\text{LUT}}(\boldsymbol{N}(x,y) \cdot \omega_o, R(x,y)) \cdot \\
&\left( V(x,y) \boldsymbol{L}_{\text{dir}}(x,y,\omega_r, R(x,y)) + (1 - V(x,y)) \boldsymbol{L}_{\text{ind}}(x,y) \right).
\end{split}
\end{equation}

The first term, $\text{BRDF}_{\textbf{LUT}}$, depends solely on the view angle and roughness, which is precomputed and stored in a 2D lookup texture. $\boldsymbol{L}_{\text{dir}}(x,y,\omega_r, R(x,y))$ is the direct environment lighting queried from a learnable environment map in the reflection direction $\omega_r$, using roughness $R(x,y)$ for mipmap selection. $\boldsymbol{L}_{\text{ind}}(x,y)$ is the blended indirect lighting component. We follow the baseline method~\cite{RefGaussian} in modeling inter-reflections by approximating visibility with ray tracing an extracted mesh and encoding indirect lighting with spherical harmonics. We follow prior work~\cite{RefGaussian} and use ~\cite{NVDiffRec} for the PBR shading.

\subsection{Training}

Splat parameters $\boldsymbol{p}_k, \boldsymbol{t}_{u_k}, \boldsymbol{t}_{v_k}, s_{u_k}, s_{v_k}, o_k$, their material texture maps ${\mathcal{T}_k^{\boldsymbol{\lambda}}, \mathcal{T}_k^{m}, \mathcal{T}_k^{r}, \mathcal{T}_k^{\boldsymbol{n}^t}}$, per-splat indirect lighting SH coefficients for $\boldsymbol{l}^{\text{ind}}_k$, and the environment maps $(\mathcal{L}_{\text{env}}^{\text{diffuse}}, \mathcal{L}_{\text{env}}^{\text{spec}})$, are optimized end-to-end using a composite loss:
\begin{equation} \mathcal{L} = \mathcal{L}_{\text{img}} + \lambda_\text{n} \mathcal{L}_\text{n}
, 
\end{equation}
where $\mathcal{L}_{\text{img}} = (1 - \lambda) \mathcal{L}_1 + \lambda \mathcal{L}_{\text{D-SSIM}}$ is the RGB reconstruction loss with balancing weight $\lambda=0.2$. The normal consistency loss $\mathcal{L}_n = 1 - \tilde{\boldsymbol{N}}(x,y)^\mathrm{T} \boldsymbol{N}(x,y)$ encourages alignment of the Gaussians with the surface by minimizing the cosine difference between the rendered normal $\boldsymbol{N}(x,y)$ and the surface normal $\tilde{\boldsymbol{N}}(x,y)$ derived from rendered depth. 
We use a single NVIDIA RTX A6000 GPU in our experiments.

\section{Experiments}

\begin{figure}[h!]
    \centering
    \includegraphics[width=\columnwidth]{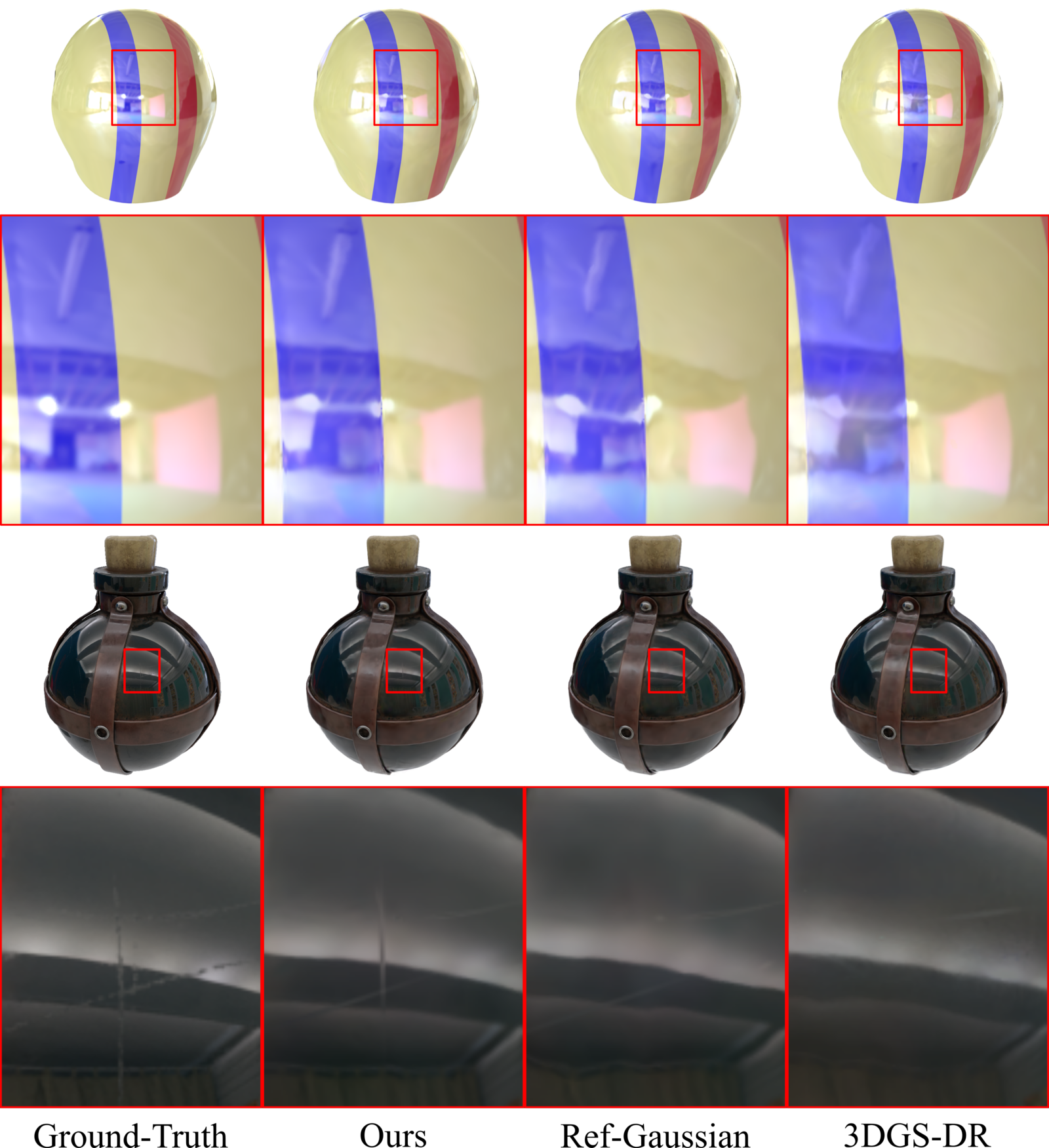}
    \caption{Qualitative comparisons of novel view synthesis on synthetic scenes. From top to bottom: helmet from Shiny Blender~\cite{ref_nerf} and potion from Glossy Synthetic~\cite{nero}. Notice how we reconstruct reflections with more fidelity and less distortion.}
    \label{fig:nvs_synth}
\end{figure}

\begin{table*}
\caption{Per-scene image quality comparison in the reflective novel view synthesis setting.
\label{tab:NVS_ref}
}
\tabcolsep=0.10cm
\renewcommand\arraystretch{1.2}
\resizebox{\linewidth}{!}{
\begin{tabular}{cc|cccccc|cccccccc|ccc}
    \hline
    \multicolumn{2}{c|}{}                                         & \multicolumn{6}{c|}{Shiny Blender~\cite{ref_nerf}}                                                                                                                                                                                                                                                                                                    & \multicolumn{8}{c|}{Glossy Synthetic~\cite{nero}}                                                                                                                                                                                                                                                                                                   & \multicolumn{3}{c}{Real~\cite{ref_nerf}}                                                                                                                                           \\ \cline{3-19} 
    \multicolumn{2}{c|}{\multirow{-2}{*}{Datasets}}               & ball    & car       & coffee    & helmet    & teapot                                               & toaster   & angel                                           & bell                                                 & cat                        & horse                          & luyu                                                 & potion                                               & tbell                                                & teapot                                               & garden                                               & sedan                                                & toycar                                               \\ \hline
    \multicolumn{1}{c|}{}                     & Ref-NeRF          & 33.16   & 30.44     & 33.99     & 29.94     & 45.12     & 26.12     & 20.89     & 30.02     & 29.76     & 19.30     & 25.42     & 30.11     & 26.91     & 22.77     & 22.01     & 25.21     & 23.65    \\
    \multicolumn{1}{c|}{}                     & ENVIDR            & \cellcolor{c1} 41.02  & 27.81     & 30.57    & \cellcolor{c2}32.71    & 42.62     & 26.03     & 29.02     & \cellcolor{c1} 30.88    & 31.04     & 25.99     & 28.03     & 32.11     & 28.64    & \cellcolor{c1}26.77     & 21.47     & 24.61     & 22.92                                                \\
    \multicolumn{1}{c|}{}                     & 3DGS              & 27.65   & 27.26     & 32.30     & 28.22     & 45.71     & 20.99     & 24.49     & 25.11     & 31.36     & 24.63     & 26.97     & 30.16     & 23.88     & 21.51     & 21.75     & 26.03     & 23.78    \\
    \multicolumn{1}{c|}{}                     & 2DGS              & 25.97   & 26.38     & 32.31     & 27.42     & 44.97     & 20.42     & 26.95     & 24.79     & 30.65     & 25.18     & 26.89     & 29.50     & 23.28     & 21.29     & 22.53     & 26.23     & 23.70                                                \\    
    \multicolumn{1}{c|}{}                     & GShader           & 30.99   & 27.96     & 32.39     & 28.32     & 45.86     & 26.28     & 25.08     & 28.07     & 31.81     & 26.56     & 27.18     & 30.09     & 24.48     & 23.58     & 21.74     & 24.89     & 23.76                                                \\
    \multicolumn{1}{c|}{}                     & 3DGS-DR           & 33.43   & 30.48    & \cellcolor{c2}34.53    & 31.44     & 47.04     & 26.76     & 29.07    & \cellcolor{c2}30.60    & 32.59     & 26.17     & 28.96     & 32.65     & 29.03     & 25.77     & 21.82    & \cellcolor{c2}26.32     & 23.83    \\
    \cline{2-19} \multicolumn{1}{c|}{}                     & Ref-Gaussian      & 36.07  & \cellcolor{c2}31.32    & 34.2      & 32.3      & \cellcolor{c2}47.15   & \cellcolor{c2}28.28   & \cellcolor{c2}30.55    & 28.57    & \cellcolor{c2}33.04    & \cellcolor{c2}26.76    & \cellcolor{c2}30.1     & \cellcolor{c2}33.39    & \cellcolor{c2}30.1      & 25.97    & \cellcolor{c2}23.09     & 26.23    & \cellcolor{c2}24.74    \\
    \multicolumn{1}{c|}{\multirow{-8}{*}{PSNR $\uparrow$}} & Ours& \cellcolor{c2}39.27  & \cellcolor{c1} 31.72     & \cellcolor{c1} 34.82   & \cellcolor{c1} 32.97    & \cellcolor{c1} 48.27    & \cellcolor{c1} 28.4    & \cellcolor{c1} 30.85    & 29.16     & \cellcolor{c1} 33.51     & \cellcolor{c1} 27.08     & \cellcolor{c1} 30.48     & \cellcolor{c1} 34.03     & \cellcolor{c1} 30.77     & \cellcolor{c2} 26.42     & \cellcolor{c1} 23.34     & \cellcolor{c1} 26.45      & \cellcolor{c1} 24.95     \\ \hline

    \multicolumn{1}{c|}{}                        & Ref-NeRF          & 0.971    & 0.950     & 0.972     & 0.954     & 0.995     & 0.921     & 0.853     & 0.941     & 0.944     & 0.820     & 0.901     & 0.933     & 0.947     & 0.897     & 0.584     & 0.720     & 0.633 \\
    \multicolumn{1}{c|}{}                        & ENVIDR            & \cellcolor{c1} 0.997   & 0.943     & 0.962     & \cellcolor{c1} 0.987    & 0.995     & 0.922     & 0.934    & \cellcolor{c2}0.954     & 0.965     & 0.925     & 0.931     & 0.960     & 0.947     & \cellcolor{c1} 0.957     & 0.561     & 0.707     & 0.549                                                \\
    \multicolumn{1}{c|}{}                        & 3DGS              & 0.937    & 0.931     & 0.972     & 0.951     & 0.996     & 0.894     & 0.792     & 0.908     & 0.959     & 0.797     & 0.916     & 0.938     & 0.900     & 0.881     & 0.571     & 0.771     & 0.637    \\
    \multicolumn{1}{c|}{}                        & 2DGS              & 0.934    & 0.930     & 0.972     & 0.953    & \cellcolor{c2}0.997    & 0.892     & 0.918     & 0.911     & 0.958     & 0.909     & 0.918     & 0.939     & 0.902     & 0.886     & 0.609     & \cellcolor{c1} 0.778     & 0.597    \\
    \multicolumn{1}{c|}{}                        & GShader           & 0.966    & 0.932     & 0.971     & 0.951     & 0.996     & 0.929     & 0.914     & 0.919     & 0.961     & 0.933     & 0.914     & 0.936     & 0.898     & 0.901     & 0.576     & 0.728     & 0.637    \\
    \multicolumn{1}{c|}{}                        & 3DGS-DR           & 0.979    & 0.963    & \cellcolor{c2}0.976    & 0.971    & \cellcolor{c2}0.997    & 0.942     & 0.942     & \cellcolor{c1} 0.959     & 0.973     & 0.933     & 0.943     & 0.959    & \cellcolor{c2}0.958     & 0.942     & 0.581    & \cellcolor{c2}0.773     & 0.639 \\
    \cline{2-19} \multicolumn{1}{c|}{}                        & Ref-Gaussian      & 0.985   & \cellcolor{c2}0.966   & \cellcolor{c2}0.976    & 0.971    & \cellcolor{c2}0.997   & \cellcolor{c2}0.952   & \cellcolor{c2}0.956    & 0.943     &\ 0.975    & \cellcolor{c2}0.942    & \cellcolor{c2}0.953    & \cellcolor{c2}0.966     & 0.947     & 0.942    & \cellcolor{c2}0.628     & 0.766    & \cellcolor{c2}0.679    \\
    \multicolumn{1}{c|}{\multirow{-8}{*}{SSIM $\uparrow$}} & Ours   & \cellcolor{c2}0.992   & \cellcolor{c1} 0.969    & \cellcolor{c1} 0.977   & \cellcolor{c2}0.975    & \cellcolor{c1} 0.998    & \cellcolor{c1} 0.954    & \cellcolor{c1} 0.958    & 0.947     & \cellcolor{c1} 0.977     & \cellcolor{c1} 0.946     & \cellcolor{c1} 0.956     & \cellcolor{c1} 0.97     & \cellcolor{c1} 0.97     & \cellcolor{c2}0.947     & \cellcolor{c1} 0.631     & 0.772      & \cellcolor{c1} 0.688 \\ \hline

    \multicolumn{1}{c|}{}                        & Ref-NeRF          & 0.166    & 0.050     & 0.082     & 0.086     & 0.012     & 0.083     & 0.144     & 0.102     & 0.104     & 0.155     & 0.098     & 0.084     & 0.114     & 0.098     & 0.251     & 0.234     & \cellcolor{c1} 0.231    \\
    \multicolumn{1}{c|}{}                        & ENVIDR            & \cellcolor{c1} 0.020   & 0.046     & 0.083     & \cellcolor{c1} 0.036    & 0.009     & 0.081     & 0.067    & \cellcolor{c2}0.054     & 0.049     & 0.065     & 0.059     & 0.072     & 0.069     & \cellcolor{c1} 0.041     & 0.263     & 0.387     & 0.345                                                \\
    \multicolumn{1}{c|}{}                        & 3DGS              & 0.162    & 0.047     & 0.079     & 0.081     & 0.008     & 0.125     & 0.088     & 0.104     & 0.062     & 0.077     & 0.064     & 0.093     & 0.102     & 0.125    & \cellcolor{c2}0.248     & \cellcolor{c1} 0.206    & \cellcolor{c2}0.237                                                \\
    \multicolumn{1}{c|}{}                        & 2DGS              & 0.156    & 0.052     & 0.079     & 0.079     & 0.008     & 0.127     & 0.072     & 0.109     & 0.060     & 0.071     & 0.066     & 0.097     & 0.125     & 0.101     & 0.254     & 0.225     & 0.396                                                \\
    \multicolumn{1}{c|}{}                        & GShader           & 0.121    & 0.044     & 0.078     & 0.074     & 0.007     & 0.079     & 0.082     & 0.098     & 0.056     & 0.562     & 0.064     & 0.088     & 0.091     & 0.122     & 0.274     & 0.259     & 0.239 \\
    \multicolumn{1}{c|}{}                        & 3DGS-DR           & 0.105    & 0.033    & \cellcolor{c2}0.076    & 0.050    & \cellcolor{c2}0.006    & 0.082     & 0.052     & \cellcolor{c1} 0.050     & 0.042     & 0.057     & 0.048     & 0.068    & \cellcolor{c2}0.059     & 0.060     & \cellcolor{c1} 0.247    & \cellcolor{c2}0.208     & \cellcolor{c1} 0.231 \\
    \cline{2-19} \multicolumn{1}{c|}{}                        & Ref-Gaussian      & 0.089   & \cellcolor{c2}0.031    & 0.078     & 0.048    & \cellcolor{c2}0.006   & \cellcolor{c2}0.067   & \cellcolor{c2}0.040    & 0.067    & \cellcolor{c2}0.037    & \cellcolor{c2}0.049    & \cellcolor{c2}0.043    & \cellcolor{c2}0.061     & 0.070     & 0.059     & 0.266     & 0.258     & 0.257  \\
    \multicolumn{1}{c|}{\multirow{-8}{*}{LPIPS $\downarrow$}} & Ours& \cellcolor{c2}0.074   & \cellcolor{c1} 0.028    & \cellcolor{c1} 0.075   & \cellcolor{c2}0.042    & \cellcolor{c1} 0.004    & \cellcolor{c1} 0.063    & \cellcolor{c1} 0.038    & 0.064     & \cellcolor{c1} 0.034     & \cellcolor{c1} 0.045     & \cellcolor{c1} 0.040     & \cellcolor{c1} 0.055     & \cellcolor{c1} 0.041    & \cellcolor{c2}0.055     & 0.283     & 0.257     & 0.27 \\ \hline
    \end{tabular}
}
\end{table*}

\begin{figure}[h!]
    \centering
    \includegraphics[width=\columnwidth]{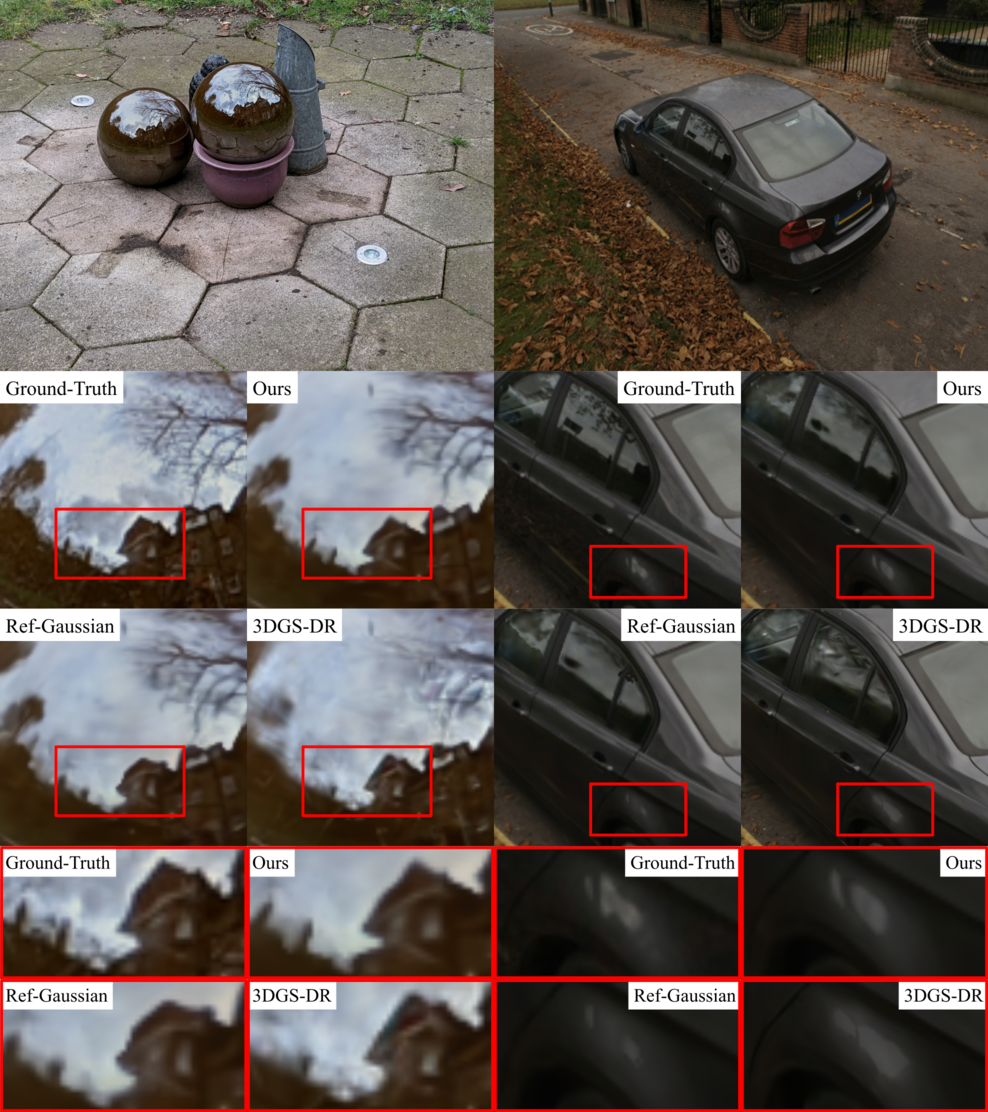}
    \caption{Qualitative comparisons of novel view synthesis on real scenes~\cite{ref_nerf}. From left to right: garden spheres and sedan. Notice how we recover reflections with more fidelity.}
    \label{fig:nvs_real}
\end{figure}

Following the baseline method~\cite{RefGaussian},
we evaluate our work quantitatively and qualitatively under standard multi-view reconstruction benchmarks of challenging reflective scenes. We use the datasets: Shiny Blender ~\cite{ref_nerf} and Glossy Synthetic~\cite{nero} for novel view synthesis of reflective objects, and dataset Ref-Real~\cite{ref_nerf} to account for real world open reflective scenes. We also provide results on the Synthetic NeRF~\cite{nerf} dataset to showcase our method under non-reflective scenes and illustrate its practicality in the supplementary material (Section 2). We compare to state-of-the-art methods in the reflective scene setting, including the baseline Ref-Gaussian~\cite{RefGaussian}, other reflective Gaussian splatting based methods GShader~\cite{Gshader} and 3DGS-DR~\cite{3DGS-DR}, 3DGS~\cite{3dgs} and 2DGS~\cite{2DGS} for reference, and seminal NeRF based approaches such as Ref-NeRF~\cite{ref_nerf} and ENVIDR~\cite{envidr}. We provide additional results and ablation studies in the supplementary material. 

\subsection{Implementation Details}
We follow a two-stage optimization for stability. We first train for half the total number of iterations using per-splat single attribute optimization. During the second stage, we start optimizing the material and normal textures initialized from the corresponding attributes from the first stage, and we freeze the positions of the primitives. In all our evaluations, we use texture resolution of $2\times2$. We use the same hyperparameters and training strategies defined by our baseline Ref-Gaussian~\cite{RefGaussian}.
We also follow the latter in replacing the integrated diffuse lighting by a spherical harmonics view dependent color for better fitting in the reflective setting. We implement efficient CUDA kernels on top of 2DGS and Ref-Gaussian for forward and backward operations involving material textures and normal mapping, as well as the texture atlases construction, packing and hardware bilinear filtering at test time.

\subsection{Novel View Synthesis}

Table \ref{tab:NVS_ref} shows numerical results in the standard reflective benchmark. 
We report Peak Signal-to-Noise Ratio (PSNR), Structural Similarity Index Measure (SSIM) \cite{wang2004image}, 
and Learned Perceptual Image Patch Similarity (LPIPS) \cite{zhang2018unreasonable} as metrics.
Our method performs favorably compared to other methods, including baseline Ref-Gaussian, which is also the state-of-the-art method currently under this benchmark to the best of our knowledge.
We provide qualitative results to accompany this table, for real scenes in Figure~\ref{fig:nvs_real} and synthetic ones in Figure~\ref{fig:nvs_synth}. Our method displays superior ability in capturing reflections on the surface with more fidelity,  while Ref-Gaussian and 3DGS-DR suffer from distorted or missing reflections.
We also show in the supplementary material numerical (Tab. 1 in Supp. Mat.) and qualitative (Fig. 1 in Supp. Mat.) comparisons under non-reflective data on Nerf Synthetic Scenes~\cite{nerf}. In that generic setting, we perform competitively with the state-of-the-art. The improvement brought by our method over the baseline Ref-Gaussian particularly, under both  reflective and non-reflective scenes, is a testimony of the efficacy and versatility of our representation. 

\subsection{Scene Decomposition} 

\begin{figure}[h!]
    \centering
    \includegraphics[width=\columnwidth]{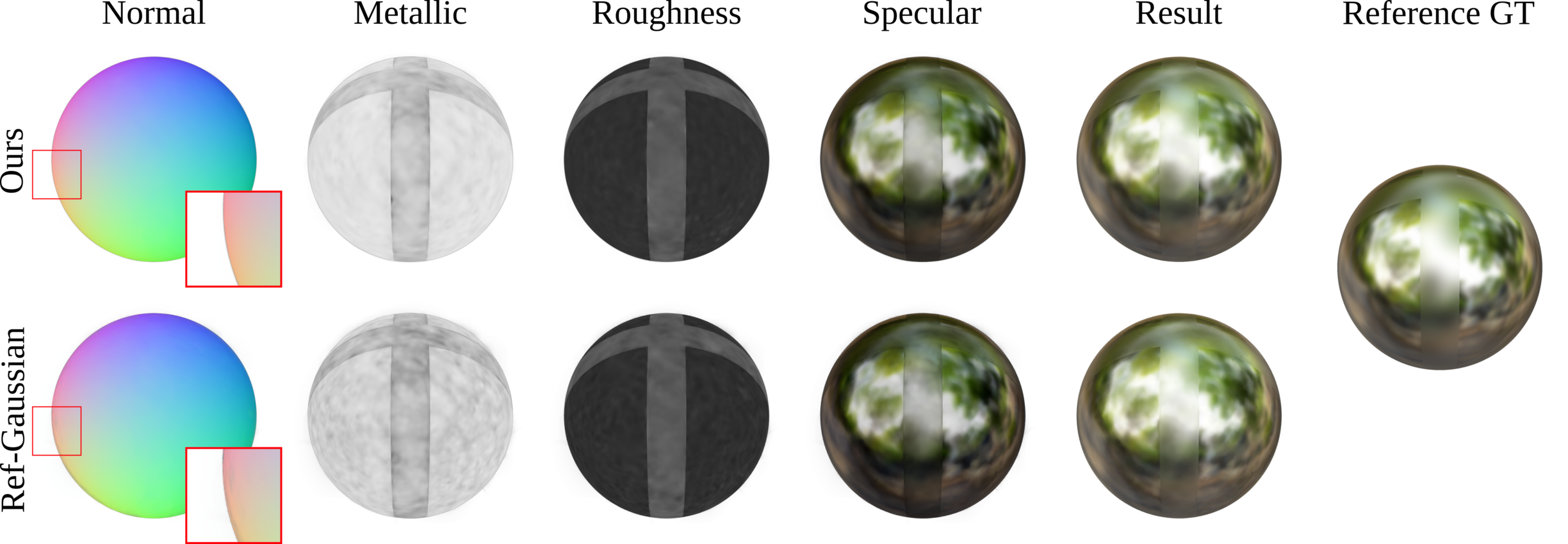}
    \caption{Comparison of scene decomposition between our method and the baseline.}
    \label{fig:decomp_compare}
\end{figure}

Figure~\ref{fig:decomp_compare} shows a comparison of the decomposition with respect to our baseline. Notice that our material properties are sharper and display less noise. Our normals replicate the smooth sphere shape more faithfully. 
We provide further comparisons in supplementary material including environment map estimation and material decomposition (Figures 3 \& 2 in Supp. Mat.) showcasing the superiority of our results.  

\begin{figure}[h!]
    \centering
    \includegraphics[width=\columnwidth]{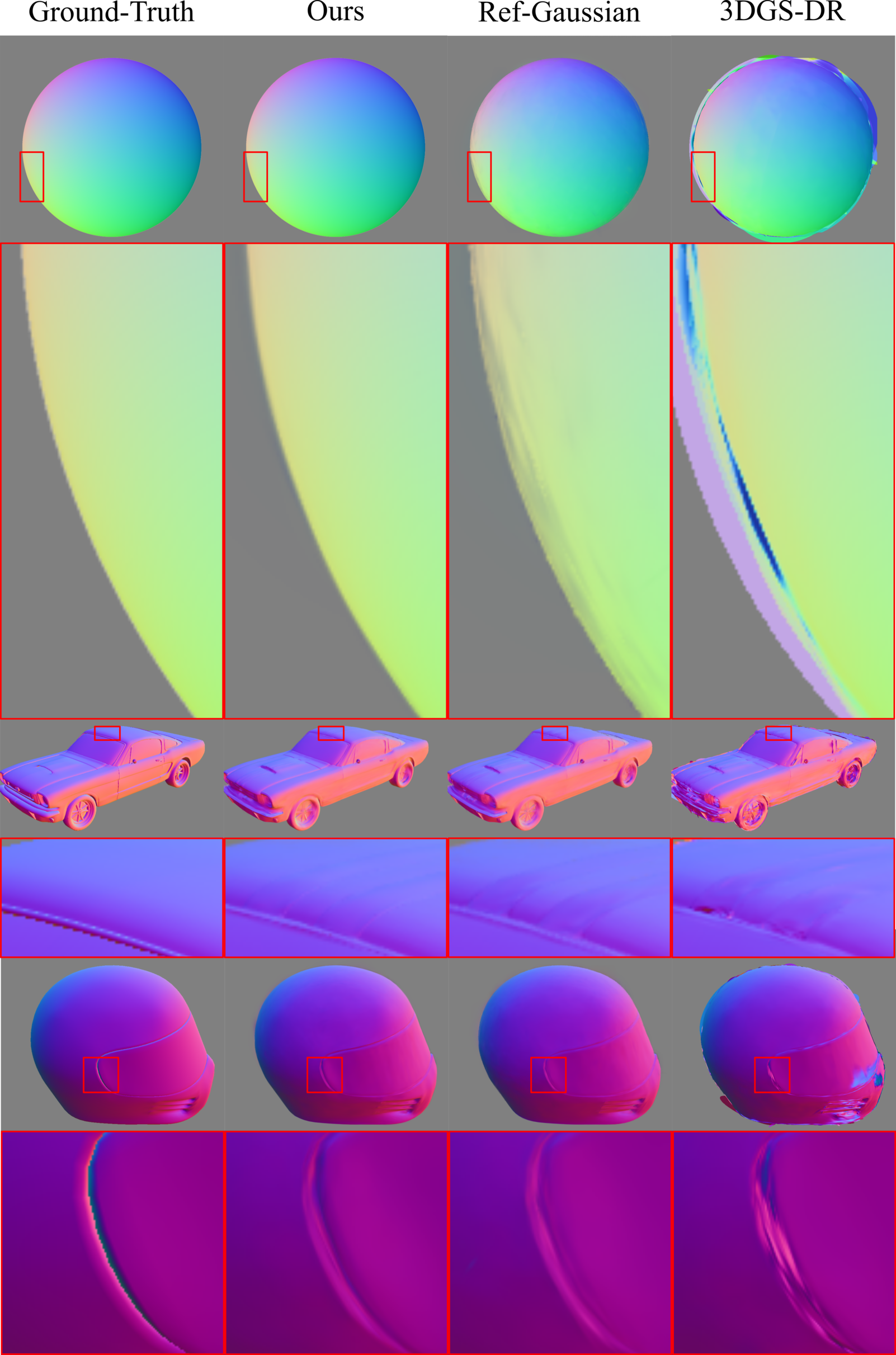}
    \caption{Qualitative comparisons of normal reconstruction by different methods.}
    \label{fig:norm_comp}
\end{figure}

We further evaluate our normal estimation through the benchmark of the Shiny Blender dataset~\cite{ref_nerf}. Table~\ref{tab:recon} reports the mean angular
error of normal maps, where we outperform our baseline. This result validates our tangential normal representation and the use of normal mapping. Figure \ref{fig:norm_comp} shows qualitative comparisons of normals, where we recover superior geometry compared to other methods.  

\begin{table}
\caption{Normal quality evaluated by MAE$^\circ$: comparisons on the Shiny Blender Dataset~\cite{ref_nerf}.}
\label{tab:recon}
\resizebox{1.0\columnwidth}{!}{
\begin{tabular}{c||ccccccc}
    \hline
          & GShader & NVDiffRec & ENVIDR & 3DGS-DR & Ref-Gaussian & Ours \\ \hline
    MAE$^\circ$ $\downarrow$   & 22.31   & 17.02     & 4.618  & 4.871 & 2.078\cellcolor{c2} & 1.78\cellcolor{c1}  \\
    \hline
\end{tabular}
}
\end{table}

\subsection{Hardware Acceleration Performance}

While our per-primitive texture mapping approach improves rendering quality for reflective scenes, a potential concern is the additional computational cost of texture sampling. To address this, we implemented hardware-accelerated texture filtering using texture atlases as described in Section~\ref{sec:hardware_acceleration}. In this section, we evaluate the performance benefits of this implementation compared to software-based bilinear filtering.

We trained our model on the Shiny Blender dataset using different texture resolutions: 2×2, 4×4, 8×8 and 16×16. For each model, we then rendered the scenes using both the software implementation of bilinear filtering and our hardware-accelerated implementation with texture atlases. We measured the rendering performance in frames per second (FPS) and compared these values to the baseline method without textures.

\begin{table}[t]
\centering
\caption{Rendering performance comparison between software bilinear filtering and hardware-accelerated texture atlas filtering at different texture resolutions. Values represent the ratio of FPS compared to the baseline method without textures, averaged across the Shiny Blender dataset scenes~\cite{ref_nerf}.}
\label{tab:hw_acceleration}
\resizebox{1.0\columnwidth}{!}{
\begin{tabular}{lcccc}
\toprule
& 2×2 Textures & 4×4 Textures & 8×8 Textures & 16×16 Textures \\
\midrule

Baseline, No Textures & $\times$ 1.00 &  $\times$ 1.00 &  $\times$ 1.00 &  $\times$ 1.00 \\
Software Bilinear &  $\times$ 0.90 &  $\times$ 0.86 &  $\times$ 0.85 &  $\times$ 0.79 \\
Hardware Texture Atlas &  $\times$ 0.92 &  $\times$ 0.93 &  $\times$ 0.94 &  $\times$ 0.91 \\
\bottomrule
\end{tabular}
}
\end{table}

As shown in Table~\ref{tab:hw_acceleration}, software-based bilinear filtering introduces some performance overhead compared to the baseline. This overhead increases with higher texture resolutions. In contrast, our hardware-accelerated implementation with texture atlases maintains rendering performance very close to the baseline method in comparison.

These results demonstrate that the texture atlas approach effectively leverages GPU hardware capabilities to minimize the performance impact of texture filtering. Several factors contribute to this efficiency including the use of dedicated texture units that are designed specifically for texture filtering operations, and are substantially faster than general-purpose compute.

We note that there is still room for improvement as we do not follow any strategy for packing the texture into the atlases, which could benefit from better locality if considering only primitives that are used for each frame or packing textures belonging to nearby primitives next to each other.

By leveraging hardware acceleration through texture atlases, we can achieve the best of both worlds: the improved rendering quality of per-primitive textures for reflective scenes while maintaining rendering speed comparable to methods without textures.

\subsection{Ablation Study}
To evaluate the effectiveness of our per-primitive texture mapping approach and understand the contribution of individual components, we conducted a series of ablation studies isolating  different aspects of our method.

Additional experiments can be found in Sec. 3 of the supplementary material, notably a comparative analysis of Texture Mapping as opposed to increasing primitive count, and a comparative analysis of performance under reduced primitive count (Tab. 2 \& Fig. 4 in Supp. Mat.).  

\paragraph{Impact of Normal Mapping}

Our previous experiments suggested that normal mapping plays a particularly important role in the performance improvements observed with our method. To isolate this effect, we implemented a variant of the baseline that uses texture mapping only for normals while keeping other attributes (albedo, roughness, metallic) as per-primitive constants.

\begin{table}
\caption{Ablation study: Effect of normal mapping on average image and normal quality. We evaluate on the Shiny Blender dataset~\cite{ref_nerf}.}
\label{tab:normal_mapping}
\vspace{-10pt}
\tabcolsep=0.15cm
\renewcommand\arraystretch{1.2}
\begin{center}
\resizebox{1.0\columnwidth}{!}{
\begin{tabular}{c|ccc}
\hline
Metric  & Ref-Gaussian & + Normal Mapping & Full Model(Ours) \\ \hline
PSNR $\uparrow$  & 34.89 & \cellcolor{c2}35.33 & \cellcolor{c1}35.90 \\
SSIM $\uparrow$  & 0.974 & \cellcolor{c2}0.975 & \cellcolor{c1}0.978 \\
LPIPS $\downarrow$  & 0.053 & \cellcolor{c2}0.053 & \cellcolor{c1}0.047 \\
Normal MAE $\downarrow$  & 2.078 & \cellcolor{c2}1.783 & \cellcolor{c1}1.78 \\
\hline
\end{tabular}
}
\end{center}
\end{table}

\begin{table}
\caption{Ablation study: Effect of texture resolution on storage, average image and normal quality. We evaluate on the Shiny Blender dataset~\cite{ref_nerf}.}
\label{tab:storage}
\vspace{-10pt}
\tabcolsep=0.15cm
\renewcommand\arraystretch{1.2}
\begin{center}
\resizebox{1.0\columnwidth}{!}{
\begin{tabular}{c|ccccc}
\hline
Variants                     & PSNR $\uparrow$     & SSIM $\uparrow$     & LPIPS $\downarrow$  & Normal MAE $\downarrow$ & Storage \\ \hline
Ref-Gaussian                & 34.88               & 0.974               & 0.053               & 2.078                   & $\times$ 1.00      \\ \hline
2$\times$2 Textures, Same Storage & 35.66               & 0.976               & \cellcolor{c2}0.048 & \cellcolor{c2}1.823     & $\times$ 1.00      \\
 2$\times$2 Textures              & \cellcolor{c2}35.90 & \cellcolor{c2}0.977 & \cellcolor{c1}0.047 & \cellcolor{c1}1.780     & $\times$ 1.20    \\ \hline
4$\times$4 Textures, Same Storage & 35.33               & \cellcolor{c2}0.977 & \cellcolor{c2}0.048 & 1.881                   & $\times$ 1.00      \\
 4$\times$4 Textures              & \cellcolor{c1}35.94 & \cellcolor{c1}0.978 & \cellcolor{c1}0.047 & 1.830                   & $\times$ 1.96   \\ \hline
\end{tabular}
}
\end{center}
\end{table}

The results in Table~\ref{tab:normal_mapping} confirm that normal mapping alone accounts for a substantial portion of the performance improvement. This is particularly evident in the normal mean angular error (MAE) metric, where normal mapping significantly reduces the error compared to the baseline.

The effectiveness of normal mapping for reflective scenes can be attributed to the fact that reflective surfaces often exhibit high-frequency normal variations that are difficult to capture with a single normal per primitive and that more accurate and detailed normals lead to more precise reflection directions and therefore better specular highlights which is achieved by our spatially varying normals.

While normal mapping provides significant improvement, our full model with texture mapping for all material properties achieves the best overall performance, demonstrating that each component contributes to the final quality.

\paragraph{Impact of Texture Resolution and Storage}
The results in Table~\ref{tab:storage} show different configurations of our method in terms of the texture resolution, and whether we add them on top of the baseline or reduce the primitive count so that the total storage size is equal to the baseline. We perform this evaluation on the Shiny Blender dataset~\cite{ref_nerf}. We experiment with resolutions 2$\times$2 and 4$\times$4 as we find that for the type of scenes in this dataset (object at the center of the scene and orbital views), enhancing primitives with much higher texture resolutions does not result in a desirable storage to quality tradeoff. Using higher resolutions can still be beneficial though for other scene types where some views are close to the object for instance.

We find that at a fixed storage budget, our method still achieves better performance compared to the baseline method in all metrics. This demonstrates the efficiency of our texture-based representation and that the memory overhead is not a fixed cost but a tunable parameter. Our framework allows for adapting texture resolutions or selectively texturing only the most critical attributes to fit different memory budgets.

Supplementary material includes additional comparisons to the baseline with increased primitive count as well as the impact on performance with reduced primitive count; These experiments show that for reflective scenes, investing the parameter budget in appearance/normal complexity (textures) is often more effective than investing it purely in geometric complexity (primitives).

\section{Limitations and Discussion}

While our per-primitive texture mapping approach significantly enhances the representation power of 2D Gaussian Splatting for the reflective setting, it comes with some limitations that present opportunities for future work.

\paragraph{Uniform Texture Resolution}

In the current implementation, we assign textures of uniform resolution to all primitives regardless of their size or importance in the scene. This approach can be inefficient for large unbounded scenes where distant primitives occupy few pixels but still receive the same texture resolution as foreground elements. A more sophisticated approach would involve adaptive texture allocation based on primitive size, viewing distance, or local detail complexity. Primitives in the background could potentially use single attribute values while reserving detailed textures for more predominant foreground elements.

\paragraph{Filtering Limitations}

Our implementation currently relies on bilinear filtering for texture sampling. While effective for our evaluated scenes, it does not fully resolve texture minification artifacts that might manifest in more challenging scenarios. while implementing proper mipmapping with trilinear filtering could be beneficial, it would be challenging to integrate with the non-antialiased nature of 2D Gaussian primitives. Potential solutions could involve combining our approach with screen-space antialiasing techniques or developing specialized filtering methods for Gaussian-based representations.

Despite these limitations, our experimental results demonstrate that per-primitive texture mapping significantly improves the visual quality of reflective scenes while maintaining real-time rendering performance.
\section{Conclusion}
We presented a method that enhances 2D Gaussian Splatting for reflective scenes by introducing per-primitive texture mapping. By leveraging the flat nature of 2D Gaussians to define textures of material properties, our approach enables high-frequency detail representation without increasing primitive count. Our hardware-accelerated implementation using texture atlases demonstrates that classical computer graphics techniques can be effectively integrated with modern differentiable rendering approaches. The results show that this representation significantly improves the quality of specular reflections, particularly through detailed normal mapping, while maintaining real-time performance. Our work bridges the gap between explicit primitive-based representations and high-quality material modeling, offering advantages of both approaches.

\section*{Acknowledgments}
\noindent This work was granted access to the HPC resources of IDRIS under the allocation 20XX-AD010616156 made by GENCI.
{
    \small
    \bibliographystyle{ieeenat_fullname}
    \bibliography{main}

@String(CVPR= {IEEE Conf. Comput. Vis. Pattern Recog.})

@String(ICCV= {Int. Conf. Comput. Vis.})

@String(ECCV= {Eur. Conf. Comput. Vis.})

@String(TOG= {ACM Trans. Graph.})

@String(ICASSP=	{ICASSP})

@String(ICIP = {IEEE Int. Conf. Image Process.})

@String(ICLR = {Int. Conf. Learn. Represent.})

@String(CVPRW= {IEEE Conf. Comput. Vis. Pattern Recog. Worksh.})

@String(CVPR  = {CVPR})

@String(ICCV  = {ICCV})

@String(ECCV  = {ECCV})

@String(TOG   = {ACM TOG})

@String(ICIP  = {ICIP})

@String(ICLR  = {ICLR})

@String(CVPRW= {CVPRW})

@inproceedings{ref_nerf,
  title={Ref-nerf: Structured view-dependent appearance for neural radiance fields},
  author={Verbin, Dor and Hedman, Peter and Mildenhall, Ben and Zickler, Todd and Barron, Jonathan T and Srinivasan, Pratul P},
  booktitle={Proceedings of the IEEE/CVF CVPR},
  pages={5481--5490},
  year={2022},
  organization={IEEE}
}

@article{nero,
author = {Liu, Yuan and Wang, Peng and Lin, Cheng and Long, Xiaoxiao and Wang, Jiepeng and Liu, Lingjie and Komura, Taku and Wang, Wenping},
title = {NeRO: Neural Geometry and BRDF Reconstruction of Reflective Objects from Multiview Images},
year = {2023},
issue_date = {August 2023},
publisher = {Association for Computing Machinery},
address = {New York, NY, USA},
volume = {42},
number = {4},
issn = {0730-0301},
journal = {ACM Trans. Graph.},
month = {jul},
articleno = {114},
numpages = {22}
}

@inproceedings{2dgs,
  title={2d gaussian splatting for geometrically accurate radiance fields},
  author={Huang, Binbin and Yu, Zehao and Chen, Anpei and Geiger, Andreas and Gao, Shenghua},
  booktitle={ACM SIGGRAPH 2024 conference papers},
  pages={1--11},
  year={2024}
}

@article{3dgs,
  title={3d gaussian splatting for real-time radiance field rendering.},
  author={Kerbl, Bernhard and Kopanas, Georgios and Leimk{\"u}hler, Thomas and Drettakis, George},
  journal={ACM Trans. Graph.},
  volume={42},
  number={4},
  pages={139--1},
  year={2023}
}

@article{ewa,
  title={EWA splatting},
  author={Zwicker, Matthias and Pfister, Hanspeter and Van Baar, Jeroen and Gross, Markus},
  journal={IEEE Transactions on Visualization and Computer Graphics},
  volume={8},
  number={3},
  pages={223--238},
  year={2002},
  publisher={IEEE}
}

@inproceedings{RefGaussian,
  title={Reflective Gaussian Splatting},
  author={Yao, Yuxuan and Zeng, Zixuan and Gu, Chun and Zhu, Xiatian and Zhang, Li},
  booktitle={ICLR},
  year={2025}
}

@inproceedings{nerf,
  title     = {NeRF: Representing Scenes as Neural Radiance Fields for View Synthesis},
  author    = {Mildenhall, Ben and Srinivasan, Pratul P. and Tancik, Matthew and Barron, Jonathan T. and Ramamoorthi, Ravi and Ng, Ren},
  booktitle = {ECCV},
  year      = {2020},
  pages     = {405--421},
  publisher = {Springer}
}

@article{levoy1988volume,
  title     = {Display of Surfaces from Volume Data},
  author    = {Levoy, Marc},
  journal   = {IEEE Computer Graphics and Applications},
  volume    = {8},
  number    = {3},
  pages     = {29--37},
  year      = {1988},
  publisher = {IEEE}
}

@inproceedings{jiang2024rethinking,
  title     = {Rethinking Directional Parameterization in Neural Implicit Surface Reconstruction},
  author    = {Jiang, Zijie and Xu, Tianhan and Kato, Hiroharu},
  booktitle = {Proceedings of the European Conference on Computer Vision (ECCV)},
  year      = {2024},
  pages     = {127--142},
  publisher = {Springer}
}

@inproceedings{Ref-NeuS,
  title     = {Ref-NeuS: Ambiguity-Reduced Neural Implicit Surface Learning for Multi-View Reconstruction with Reflection},
  author    = {Ge, Wenhang and Hu, Tao and Zhao, Haoyu and Liu, Shu and Chen, Ying-Cong},
  booktitle = {Proceedings of the IEEE/CVF International Conference on Computer Vision (ICCV)},
  year      = {2023},
  pages     = {4251--4260},
  publisher = {IEEE}
}

@article{neus,
  title={Neus: Learning neural implicit surfaces by volume rendering for multi-view reconstruction},
  author={Wang, Peng and Liu, Lingjie and Liu, Yuan and Theobalt, Christian and Komura, Taku and Wang, Wenping},
  journal={arXiv preprint arXiv:2106.10689},
  year={2021}
}

@article{volsdf,
  title={Volume rendering of neural implicit surfaces},
  author={Yariv, Lior and Gu, Jiatao and Kasten, Yoni and Lipman, Yaron},
  journal={Advances in Neural Information Processing Systems},
  volume={34},
  pages={4805--4815},
  year={2021}
}

@inproceedings{neuralangelo,
  title={Neuralangelo: High-fidelity neural surface reconstruction},
  author={Li, Zhaoshuo and M{\"u}ller, Thomas and Evans, Alex and Taylor, Russell H and Unberath, Mathias and Liu, Ming-Yu and Lin, Chen-Hsuan},
  booktitle={Proceedings of the IEEE/CVF Conference on Computer Vision and Pattern Recognition},
  pages={8456--8465},
  year={2023}
}

@article{ingp,
  title={Instant neural graphics primitives with a multiresolution hash encoding},
  author={M{\"u}ller, Thomas and Evans, Alex and Schied, Christoph and Keller, Alexander},
  journal={ACM transactions on graphics (TOG)},
  volume={41},
  number={4},
  pages={1--15},
  year={2022},
  publisher={ACM New York, NY, USA}
}

@inproceedings{plenoxels,
  title={Plenoxels: Radiance fields without neural networks},
  author={Fridovich-Keil, Sara and Yu, Alex and Tancik, Matthew and Chen, Qinhong and Recht, Benjamin and Kanazawa, Angjoo},
  booktitle={Proceedings of the IEEE/CVF conference on computer vision and pattern recognition},
  pages={5501--5510},
  year={2022}
}

@article{voxurf,
  title={Voxurf: Voxel-based efficient and accurate neural surface reconstruction},
  author={Wu, Tong and Wang, Jiaqi and Pan, Xingang and Xu, Xudong and Theobalt, Christian and Liu, Ziwei and Lin, Dahua},
  journal={arXiv preprint arXiv:2208.12697},
  year={2022}
}

@article{volrend,
  title={Optical models for direct volume rendering},
  author={Nelson Max},
  journal={IEEE Transactions on Visualization and Computer Graphics},
  volume={1},
  number={2},
  pages={99--108},
  year={1995},
  publisher={IEEE}
}

@inproceedings{tensorrf,
  title={TensoRF: Tensorial Radiance Fields},
  author={Anpei Chen and Zexiang Xu and Andreas Geiger and Jingyi Yu and Hao Su},
  booktitle={Proceedings of the European Conference on Computer Vision (ECCV)},
  year={2022}
}

@misc{dvgo,
  title={Improved Direct Voxel Grid Optimization for Radiance Fields},
  author={Jingxiang Sun and Yiming Gao and Xuan Wang and Qi Zhang and Hujun Bao and Xiaowei Zhou},
  year={2022},
  eprint={2206.05085},
  archivePrefix={arXiv},
  primaryClass={cs.CV}
}

@inproceedings{kplanes,
  title={K-Planes: Explicit Radiance Fields in Space, Time, and Appearance},
  author={Sara Fridovich-Keil and Giacomo Meanti and Frederik Rahbæk Warburg and Benjamin Recht and Angjoo Kanazawa},
  booktitle={Proceedings of the IEEE/CVF Conference on Computer Vision and Pattern Recognition (CVPR)},
  year={2023}
}

@inproceedings{nsvf,
  title={Neural Sparse Voxel Fields},
  author={Lingjie Liu and Jiatao Gu and Kyaw Zaw Lin and Tat-Seng Chua and Christian Theobalt},
  booktitle={NeurIPS},
  year={2020}
}

@article{sun2024sparse,
  title   = {Sparse Voxels Rasterization: Real-time High-fidelity Radiance Field Rendering},
  author  = {Sun, Cheng and Choe, Jaesung and Loop, Charles and Ma, Wei-Chiu and Wang, Yu-Chiang Frank},
  journal = {arXiv preprint arXiv:2412.04459},
  year    = {2024}
}

@inproceedings{zwicker2001ewa,
  title     = {EWA Volume Splatting},
  author    = {Zwicker, Matthias and Pfister, Hanspeter and Van Baar, Jeroen and Gross, Markus},
  booktitle = {Proceedings of the IEEE Conference on Visualization (VIS)},
  year      = {2001},
  pages     = {29--36},
  publisher = {IEEE}
}

@inproceedings{nerv,
  title     = {NeRV: Neural Reflectance and Visibility Fields for Relighting and View Synthesis},
  author    = {Srinivasan, Pratul P. and Deng, Boyang and Zhang, Xiuming and Tancik, Matthew and Mildenhall, Ben and Barron, Jonathan T.},
  booktitle = {Proceedings of the IEEE/CVF Conference on Computer Vision and Pattern Recognition (CVPR)},
  year      = {2021},
  pages     = {7495--7504},
  doi       = {10.1109/CVPR46437.2021.00741}
}

@inproceedings{envidr,
  title     = {ENVIDR: Implicit Differentiable Renderer with Neural Environment Lighting},
  author    = {Liang, Ruofan and Chen, Huiting and Li, Chunlin and Chen, Fan and Panneer, Selvakumar and Vijaykumar, Nandita},
  booktitle = {Proceedings of the IEEE/CVF International Conference on Computer Vision (ICCV)},
  year      = {2023},
  pages     = {79--89},
  publisher = {IEEE}
}

@inproceedings{Gshader,
  title     = {GaussianShader: 3D Gaussian Splatting with Shading Functions for Reflective Surfaces},
  author    = {Jiang, Yingwenqi and Tu, Jiadong and Liu, Yuan and Gao, Xifeng and Long, Xiaoxiao and Wang, Wenping and Ma, Yuexin},
  booktitle = {Proceedings of the IEEE/CVF Conference on Computer Vision and Pattern Recognition (CVPR)},
  year      = {2024},
  pages     = {1185--1194},
  publisher = {IEEE},
  doi       = {10.1109/CVPR.2024.00127}
}

@inproceedings{3DGS-DR,
  title     = {3D Gaussian Splatting with Deferred Reflection},
  author    = {Ye, Keyang and Hou, Qiming and Zhou, Kun},
  booktitle = {ACM SIGGRAPH Conference Proceedings},
  year      = {2024},
  doi       = {10.1145/3641519.3657456}
}

@inproceedings{R3DG,
  title     = {Relightable 3D Gaussians: Realistic Point Cloud Relighting with BRDF Decomposition and Ray Tracing},
  author    = {Gao, Jian and Gu, Chun and Lin, Youtian and Li, Zhihao and Zhu, Hao and Cao, Xun and Zhang, Li and Yao, Yao},
  booktitle = {Proceedings of the European Conference on Computer Vision (ECCV)},
  year      = {2024},
  url       = {https://arxiv.org/abs/2311.16043}
}

@inproceedings{GS-IR,
  author    = {Liang, Zhihao and Zhang, Qi and Feng, Ying and Shan, Ying and Jia, Kui},
  title     = {GS-IR: 3D Gaussian Splatting for Inverse Rendering},
  booktitle = {Proceedings of the IEEE/CVF Conference on Computer Vision and Pattern Recognition (CVPR)},
  month     = {June},
  year      = {2024},
  pages     = {21644--21653}
}

@inproceedings{NVDiffRec,
    author    = {Munkberg, Jacob and Hasselgren, Jon and Shen, Tianchang and Gao, Jun and Chen, Wenzheng 
                    and Evans, Alex and M\"uller, Thomas and Fidler, Sanja},
    title     = "{Extracting Triangular 3D Models, Materials, and Lighting From Images}",
    booktitle = {Proceedings of the IEEE/CVF Conference on Computer Vision and Pattern Recognition (CVPR)},
    month     = {June},
    year      = {2022},
    pages     = {8280-8290}
}

@article{GS-ROR,
  title   = {GS-ROR: 3D Gaussian Splatting for Reflective Object Relighting via SDF Priors},
  author  = {Zhu, Zuo-Liang and Wang, Beibei and Yang, Jian},
  journal = {arXiv preprint arXiv:2406.18544},
  year    = {2024},
  url     = {https://arxiv.org/abs/2406.18544}
}

@inproceedings{3iGS,
  title     = {3iGS: Factorised Tensorial Illumination for 3D Gaussian Splatting},
  author    = {Tang, Zhe Jun and Cham, Tat-Jen},
  booktitle = {Proceedings of the European Conference on Computer Vision (ECCV)},
  year      = {2024},
  publisher = {Springer},
  doi       = {10.1007/978-3-031-72630-9_9}
}

@article{phong,
  title={Illumination for computer generated pictures},
  author={Phong, Bui Tuong},
  journal={Communications of the ACM},
  volume={18},
  number={6},
  pages={311--317},
  year={1975},
  publisher={ACM}
}

@inproceedings{gouraud,
  title={Continuous shading of curved surfaces},
  author={Gouraud, Henri},
  booktitle={Proceedings of the 2nd conference on Computer graphics and interactive techniques},
  pages={747--755},
  year={1971},
  organization={ACM}
}

@article{wang2004image,
  title={Image quality assessment: from error visibility to structural similarity},
  author={Wang, Zhou and Bovik, Alan C and Sheikh, Hamid R and Simoncelli, Eero P},
  journal={IEEE transactions on image processing},
  volume={13},
  number={4},
  pages={600--612},
  year={2004},
  publisher={IEEE}
}

@inproceedings{zhang2018unreasonable,
  title={The unreasonable effectiveness of deep features as a perceptual metric},
  author={Zhang, Richard and Isola, Phillip and Efros, Alexei A and Shechtman, Eli and Wang, Oliver},
  booktitle={Proceedings of the IEEE conference on computer vision and pattern recognition},
  pages={586--595},
  year={2018}
}

@inproceedings{Burley2008Ptex,
  author    = {Brent Burley and Dylan Lacewell},
  title     = {Ptex: Per-Face Texture Mapping for Production Rendering},
  booktitle = {Proceedings of Eurographics Symposium on Rendering},
  year      = {2008},
  publisher = {Eurographics Association},
  url       = {https://graphics.pixar.com/library/Ptex/paper.pdf}
}

@inproceedings{purnomo2004seamless,
  title={Seamless texture atlases},
  author={Purnomo, Budirijanto and Cohen, Jonathan D and Kumar, Subodh},
  booktitle={IEEE Visualization, 2004. VIS 2004.},
  pages={273--280},
  year={2004},
  organization={IEEE},
  doi={10.1109/VISUAL.2004.91}
}

@InProceedings{AA-2DGS,
  author    = {Mae Younes and Adnane Boukhayma},
  title     = {Anti-Aliased 2D Gaussian Splatting},
  booktitle = {Proceedings of the 39th Conference on Neural Information Processing Systems (NeurIPS)},
  year      = {2025},
  note      = {Poster}
}

@article{GTAvatar,
  title        = {GTAvatar: Bridging Gaussian Splatting and Texture Mapping for
Relightable and Editable Gaussian Avatars},
  author       = {Kelian Baert and Mae Younes and Francois Bourel and Marc Christie and Adnane Boukhayma},
  year         = {2025},
  journal={arXiv preprint}
}

@InProceedings{SparseCraft,
  author    = {Mae Younes and Amine Ouasfi and Adnane Boukhayma},
  title     = {SparseCraft: Few-Shot Neural Reconstruction through Stereopsis Guided Geometric Linearization},
  booktitle = {Proceedings of the European Conference on Computer Vision (ECCV)},
  year      = {2024},
  publisher = {Springer}
}

@InProceedings{GeoTransfer,
  author    = {Shubhendu Jena and Franck Multon and Adnane Boukhayma},
  title     = {GeoTransfer: Generalizable Few-Shot Multi-View Reconstruction via Transfer Learning},
  booktitle = {Proceedings of the ECCV 2024 Workshop on (insert full workshop name here)},
  year      = {2024},
}

@inproceedings{dro,
  title = {Toward Robust Neural Reconstruction from Sparse Point Sets},
  author = {Amine Ouasfi and Shubhendu Jena and Eric Marchand and Adnane Boukhayma},
  booktitle = {Proceedings of the IEEE/CVF Conference on Computer Vision and Pattern Recognition (CVPR)},
  year = {2025}
}

@inproceedings{nap,
  title = {Few-Shot Unsupervised Implicit Neural Shape Representation Learning with Spatial Adversaries},
  author = {Amine Ouasfi and Adnane Boukhayma},
  booktitle = {Proceedings of the 41st International Conference on Machine Learning (ICML)},
  series = {PMLR 235},
  pages = {38905--38918},
  year = {2024}
}

@inproceedings{sparseocc,
  title={Unsupervised occupancy learning from sparse point cloud},
  author={Ouasfi, Amine and Boukhayma, Adnane},
  booktitle={Proceedings of the IEEE/CVF Conference on Computer Vision and Pattern Recognition},
  pages={21729--21739},
  year={2024}
}

@inproceedings{fssdf,
  title={Few'Zero Level Set'-Shot Learning of Shape Signed Distance Functions in Feature Space},
  author={Ouasfi, Amine and Boukhayma, Adnane},
  booktitle={ECCV},
  year={2022}
}

@inproceedings{nksr,
  title={Neural kernel surface reconstruction},
  author={Huang, Jiahui and Gojcic, Zan and Atzmon, Matan and Litany, Or and Fidler, Sanja and Williams, Francis},
  booktitle={Proceedings of the IEEE/CVF Conference on Computer Vision and Pattern Recognition},
  pages={4369--4379},
  year={2023}
}

@inproceedings{convocc,
  title={Convolutional occupancy networks},
  author={Peng, Songyou and Niemeyer, Michael and Mescheder, Lars and Pollefeys, Marc and Geiger, Andreas},
  booktitle={Computer Vision--ECCV 2020: 16th European Conference, Glasgow, UK, August 23--28, 2020, Proceedings, Part III 16},
  pages={523--540},
  year={2020},
  organization={Springer}
}

@inproceedings{mvsnerf,
  title={Mvsnerf: Fast generalizable radiance field reconstruction from multi-view stereo},
  author={Chen, Anpei and Xu, Zexiang and Zhao, Fuqiang and Zhang, Xiaoshuai and Xiang, Fanbo and Yu, Jingyi and Su, Hao},
  booktitle={Proceedings of the IEEE/CVF International Conference on Computer Vision},
  pages={14124--14133},
  year={2021}
}

@inproceedings{pixelnerf,
  title={pixelnerf: Neural radiance fields from one or few images},
  author={Yu, Alex and Ye, Vickie and Tancik, Matthew and Kanazawa, Angjoo},
  booktitle={Proceedings of the IEEE/CVF Conference on Computer Vision and Pattern Recognition},
  pages={4578--4587},
  year={2021}
}

@inproceedings{geonerf,
  title={Geonerf: Generalizing nerf with geometry priors},
  author={Johari, Mohammad Mahdi and Lepoittevin, Yann and Fleuret, Fran{\c{c}}ois},
  booktitle={Proceedings of the IEEE/CVF Conference on Computer Vision and Pattern Recognition},
  pages={18365--18375},
  year={2022}
}

@inproceedings{genlf,
  title={Learning generalizable light field networks from few images},
  author={Li, Qian and Multon, Franck and Boukhayma, Adnane},
  booktitle={ICASSP 2023-2023 IEEE International Conference on Acoustics, Speech and Signal Processing (ICASSP)},
  pages={1--5},
  year={2023},
  organization={IEEE}
}

@article{robust,
  title={Robustifying Generalizable Implicit Shape Networks with a Tunable Non-Parametric Model},
  author={Ouasfi, Amine and Boukhayma, Adnane},
  journal={Advances in Neural Information Processing Systems},
  volume={36},
  year={2024}
}

@inproceedings{digs,
  author    = {Yizhak Ben-Shabat and Chamin Hewa Koneputugodage and Stephen Gould},
  title     = {DiGS: Divergence Guided Shape Implicit Neural Representation for Unoriented Point Clouds},
  booktitle = {Proceedings of the IEEE/CVF Conference on Computer Vision and Pattern Recognition (CVPR)},
  year      = {2022},
  pages     = {19321--19330},
  doi       = {10.1109/CVPR52688.2022.01894}
}

@incollection{igr,
  author    = {Amos Gropp and Lior Yariv and Niv Haim and Matan Atzmon and Yaron Lipman},
  title     = {Implicit Geometric Regularization for Learning Shapes},
  booktitle = {Proceedings of Machine Learning and Systems (MLSys) 2020},
  year      = {2020},
  pages     = {3569--3579},
  url       = {https://proceedings.mlr.press/v119/gropp20a/gropp20a.pdf}
}

@InProceedings{regnerf,
  author    = {Michael Niemeyer and Jonathan T. Barron and Ben Mildenhall and Mehdi S. M. Sajjadi and Andreas Geiger and Noha Radwan},
  title     = {RegNeRF: Regularizing Neural Radiance Fields for View Synthesis from Sparse Inputs},
  booktitle = {Proceedings of the IEEE/CVF Conference on Computer Vision and Pattern Recognition (CVPR)},
  year      = {2022}
}

@InProceedings{freenerf,
  author    = {Jiawei Yang and Marco Pavone and Yue Wang},
  title     = {FreeNeRF: Improving Few-Shot Neural Rendering with Free Frequency Regularization},
  booktitle = {Proceedings of the IEEE/CVF Conference on Computer Vision and Pattern Recognition (CVPR)},
  year      = {2023},
  pages     = {8254--8263}
}

@InProceedings{dietnerf,
  author    = {Ajay Jain and Matthew Tancik and Pieter Abbeel},
  title     = {Putting NeRF on a Diet: Semantically Consistent Few-Shot View Synthesis},
  booktitle = {Proceedings of the IEEE/CVF International Conference on Computer Vision (ICCV)},
  month     = {October},
  year      = {2021}
}

@article{mixing,
  title={Mixing-denoising generalizable occupancy networks},
  author={Ouasfi, Amine and Boukhayma, Adnane},
  journal={3DV},
  year={2024}
}

@InProceedings{dsnerf,
  author    = {Kangle Deng and Andrew Liu and Jun-Yan Zhu and Deva Ramanan},
  title     = {Depth-Supervised NeRF: Fewer Views and Faster Training for Free},
  booktitle = {Proceedings of the IEEE/CVF Conference on Computer Vision and Pattern Recognition (CVPR)},
  month     = {June},
  year      = {2022},
  pages     = {12872--12881}
}

@InProceedings{rgbdnerf,
  author    = {Qian Li and Franck Multon and Adnane Boukhayma},
  title     = {Regularizing Neural Radiance Fields from Sparse RGB-D Inputs},
  booktitle = {Proceedings of the IEEE International Conference on Image Processing (ICIP)},
  year      = {2023},
  pages     = {2320--2324}
}

@inproceedings{pixsplat,
  title     = {pixelSplat: 3D Gaussian Splats from Image Pairs for Scalable Generalizable 3D Reconstruction},
  author    = {David Charatan and Sizhe Lester Li and Andrea Tagliasacchi and Vincent Sitzmann},
  booktitle = {Proceedings of the IEEE/CVF Conference on Computer Vision and Pattern Recognition (CVPR)},
  year      = {2024},
  pages     = {19457--19467},
  doi       = {10.1109/CVPR52733.2024.01839}
}

@inproceedings{mvsgaussian,
  title     = {MVSGaussian: Fast Generalizable Gaussian Splatting Reconstruction from Multi-View Stereo},
  author    = {Tianqi Liu and Guangcong Wang and Shoukang Hu and Liao Shen and Xinyi Ye and Yuhang Zang and Zhiguo Cao and Wei Li and Ziwei Liu},
  booktitle = {Proceedings of the European Conference on Computer Vision (ECCV)},
  year      = {2024},
  pages     = {2662--2678},
  doi       = {10.1007/978-3-031-72649-1_3},
  url       = {https://arxiv.org/abs/2405.12218}
}

@inproceedings{bad,
  title={Bad-gaussians: Bundle adjusted deblur gaussian splatting},
  author={Zhao, Lingzhe and Wang, Peng and Liu, Peidong},
  booktitle={European Conference on Computer Vision},
  pages={233--250},
  year={2024},
  organization={Springer}
}

@InProceedings{colfree,
    author    = {Fu, Yang and Liu, Sifei and Kulkarni, Amey and Kautz, Jan and Efros, Alexei A. and Wang, Xiaolong},
    title     = {COLMAP-Free 3D Gaussian Splatting},
    booktitle = {Proceedings of the IEEE/CVF Conference on Computer Vision and Pattern Recognition (CVPR)},
    month     = {June},
    year      = {2024},
    pages     = {20796--20805},
    doi       = {10.1109/CVPR52733.2024.01965},
    url       = {https://openaccess.thecvf.com/content/CVPR2024/papers/Fu_COLMAP-Free_3D_Gaussian_Splatting_CVPR_2024_paper.pdf}
}

@inproceedings{ndim,
  author    = {Stavros Diolatzis and Tobias Zirr and Alexandr Kuznetsov and Georgios Kopanas and Anton Kaplanyan},
  title     = {N-Dimensional Gaussians for Fitting of High Dimensional Functions},
  booktitle = {Proceedings of ACM SIGGRAPH (Conference Track)},
  year      = {2024},
  month     = {July},
  publisher = {ACM},
  doi       = {10.1145/3641519.3657502},
  url       = {https://dl.acm.org/doi/10.1145/3641519.3657502}
}

@inproceedings{4d,
  title={4d gaussian splatting for real-time dynamic scene rendering},
  author={Wu, Guanjun and Yi, Taoran and Fang, Jiemin and Xie, Lingxi and Zhang, Xiaopeng and Wei, Wei and Liu, Wenyu and Tian, Qi and Wang, Xinggang},
  booktitle={Proceedings of the IEEE/CVF conference on computer vision and pattern recognition},
  pages={20310--20320},
  year={2024}
}

@inproceedings{3dgsmcmc,
  title = {3D Gaussian Splatting as Markov Chain Monte Carlo},
  author = {Kheradmand, Shakiba and Rebain, Daniel and Sharma, Gopal and Sun, Weiwei and Tseng, Yang-Che and Isack, Hossam and Kar, Abhishek and Tagliasacchi, Andrea and Yi, Kwang Moo},
  booktitle = {Advances in Neural Information Processing Systems (NeurIPS)},
  year = {2024},
  note = {Spotlight Presentation},
  url = {https://proceedings.neurips.cc/paper_files/paper/2024/file/93be245fce00a9bb2333c17ceae4b732-Paper-Conference.pdf}
}

@inproceedings{compact,
  title={Compact 3d gaussian representation for radiance field},
  author={Lee, Joo Chan and Rho, Daniel and Sun, Xiangyu and Ko, Jong Hwan and Park, Eunbyung},
  booktitle={Proceedings of the IEEE/CVF Conference on Computer Vision and Pattern Recognition},
  pages={21719--21728},
  year={2024}
}

@article{newton,
  title={3DGS2: Near Second-order Converging 3D Gaussian Splatting},
  author={Lan, Lei and Shao, Tianjia and Lu, Zixuan and Zhang, Yu and Jiang, Chenfanfu and Yang, Yin},
  journal={arXiv preprint arXiv:2501.13975},
  year={2025}
}

@inproceedings{gstex,
  title={Gstex: Per-primitive texturing of 2d gaussian splatting for decoupled appearance and geometry modeling},
  author={Rong, Victor and Chen, Jingxiang and Bahmani, Sherwin and Kutulakos, Kiriakos N and Lindell, David B},
  booktitle={2025 IEEE/CVF Winter Conference on Applications of Computer Vision (WACV)},
  pages={3508--3518},
  year={2025},
  organization={IEEE}
}

@misc{texturedgaussiansenhanced3d,
      title={Textured Gaussians for Enhanced 3D Scene Appearance Modeling}, 
      author={Brian Chao and Hung-Yu Tseng and Lorenzo Porzi and Chen Gao and Tuotuo Li and Qinbo Li and Ayush Saraf and Jia-Bin Huang and Johannes Kopf and Gordon Wetzstein and Changil Kim},
      year={2025},
      eprint={2411.18625},
      archivePrefix={arXiv},
      primaryClass={cs.CV},
      url={https://arxiv.org/abs/2411.18625}, 
}

@article{BBSplat,
  title={BillBoard Splatting (BBSplat): Learnable Textured Primitives for Novel View Synthesis},
  author={Svitov, David and Morerio, Pietro and Agapito, Lourdes and Del Bue, Alessio},
  journal={arXiv preprint arXiv:2411.08508},
  year={2024}
}

@article{SuperGaussians,
  author  = {Rui Xu and Wenyue Chen and Jiepeng Wang and Yuan Liu and Peng Wang and Lin Gao and Shiqing Xin and Taku Komura and Xin Li and Wenping Wang},
  title   = {SuperGaussians: Enhancing Gaussian Splatting Using Primitives with Spatially Varying Colors},
  journal = {arXiv preprint arXiv:2411.18966},
  year    = {2024},
  url     = {https://arxiv.org/abs/2411.18966}
}

@article{GaussianBillboards,
  author  = {Sebastian Weiss and Derek Bradley},
  title   = {Gaussian Billboards: Expressive 2D Gaussian Splatting with Textures},
  journal = {arXiv preprint arXiv:2412.12734},
  year    = {2024},
  url     = {https://arxiv.org/abs/2412.12734}
}

@article{HDGS,
  author  = {Yunzhou Song and Heguang Lin and Jiahui Lei and Lingjie Liu and Kostas Daniilidis},
  title   = {HDGS: Textured 2D Gaussian Splatting for Enhanced Scene Rendering},
  journal = {arXiv preprint arXiv:2412.01823},
  year    = {2024},
  url     = {https://arxiv.org/abs/2412.01823}
}

@InProceedings{Sparfels,
  author    = {Shubhendu Jena and Amine Ouasfi and Mae Younes and Adnane Boukhayma},
  title     = {Sparfels: Fast Reconstruction from Sparse Unposed Imagery},
  booktitle = {Proceedings of the IEEE/CVF International Conference on Computer Vision (ICCV)},
  month     = {October},
  year      = {2025}  
}

@InProceedings{SparSplat,
  author    = {Shubhendu Jena and Shishir Reddy Vutukur and Adnane Boukhayma},
  title     = {SparSplat: Fast Multi-View Reconstruction with Generalizable 2D Gaussian Splatting},
  booktitle = {Proceedings of the IEEE/CVF Conference on Computer Vision and Pattern Recognition Workshops (CVPRW)},
  year      = {2025},
}

@inproceedings{HiSplat,
  title={HiSplat: Hierarchical 3D Gaussian Splatting for Generalizable Sparse-View Reconstruction},
  author={Tang, Shengji and Ye, Weicai and Ye, Peng and Lin, Weihao and Zhou, Yang and Chen, Tao and Ouyang, Wanli},
  booktitle={Proceedings of the 2025 International Conference on Learning Representations (ICLR)},
  year={2025}
}

@inproceedings{noposplat,
  title={No Pose, No Problem: Surprisingly Simple 3D Gaussian Splats from Sparse Unposed Images},
  author={Ye, Botao and Liu, Sifei and Xu, Haofei and Li, Xueting and Pollefeys, Marc and Yang, Ming-Hsuan and Peng, Songyou},
  booktitle={Proceedings of the 2025 International Conference on Learning Representations (ICLR)},
  year={2025},
  url={https://openreview.net/forum?id=P4o9akekdf}
}

@inproceedings{steepgs,
  title={Steepest Descent Density Control for Compact 3D Gaussian Splatting},
  author={Peng Wang and Zhihao Li and Jong Hwan Ko and Eunbyung Park},
  booktitle={IEEE/CVF Conference on Computer Vision and Pattern Recognition (CVPR)},
  year={2025}
}

@inproceedings{contextgs,
  title={Compact 3D Gaussian Splatting with Anchor Level Context Model},
  author={Yufei Wang and Zhihao Li and Lanqing Guo and Wenhan Yang and Alex Kot and Bihan Wen},
  booktitle={Advances in Neural Information Processing Systems (NeurIPS)},
  year={2024}
}

@inproceedings{grendel,
  title={Grendel: On Scaling Up 3D Gaussian Splatting Training},
  author={Haoyang Zhao and Zhihao Li and Yufei Wang and others},
  booktitle={International Conference on Learning Representations (ICLR)},
  year={2025}
}

@inproceedings{speedysplat,
  title={Speedy-Splat: Fast 3D Gaussian Splatting with Sparse Pixels and Sparse Gaussians},
  author={Aaron Hanson and Zhihao Li and Yufei Wang and others},
  booktitle={IEEE/CVF Conference on Computer Vision and Pattern Recognition (CVPR)},
  year={2025}
}

@inproceedings{pixelgs,
  title={Pixel-GS: Density Control with Pixel-aware Gradient for 3D Gaussian Splatting},
  author={Zheng Zhang and Wenbo Hu and Yixing Lao and Tong He and Hengshuang Zhao},
  booktitle={European Conference on Computer Vision (ECCV)},
  year={2024}
}

@inproceedings{mipsplat,
  title     = {Mip-Splatting: Alias-free 3D Gaussian Splatting},
  author    = {Zehao Yu and Anpei Chen and Binbin Huang and Torsten Sattler and Andreas Geiger},
  booktitle = {Proceedings of the IEEE/CVF Conference on Computer Vision and Pattern Recognition (CVPR)},
  year      = {2024},
  pages     = {19447--19456},
  doi       = {10.1109/CVPR52733.2024.01839}
}

@inproceedings{anal,
  title={Analytic-splatting: Anti-aliased 3d gaussian splatting via analytic integration},
  author={Liang, Zhihao and Zhang, Qi and Hu, Wenbo and Zhu, Lei and Feng, Ying and Jia, Kui},
  booktitle={European conference on computer vision},
  pages={281--297},
  year={2024},
  organization={Springer}
}
}

\clearpage
\newpage
\begin{center}
\Large\textbf{Supplementary Material}
\end{center}

\section{Implementation Details of Hardware Accelerated Texture Atlases}
\paragraph{Texture Atlas Construction}
Each splat is associated with a small, fixed-resolution texture (e.g., $T \times T$ texels, where $T$ is typically in powers of $2$). For $P$ splats and $C$ total channels for all textured attributes, this $P \times C \times T \times T$ texel data is organized into one or more large 2D atlas textures. We pack the individual primitive textures into larger texture atlases that can be efficiently accessed by the GPU by determining the maximum atlas dimensions based on GPU capabilities, calculating how many $T \times T$ textures fit along each dimension, computing the total number of textures per atlas, then determining the number of atlases needed.

To efficiently access the textures within the atlas, we create an indirection buffer that maps primitive IDs to texture chart coordinates within the atlas. This buffer stores the top-left corner coordinates $(chart\_idx\_x, chart\_idx\_y)$ of each primitive's texture in the atlas.
Since CUDA texture objects typically support up to 4 channels, multiple atlases are used if the total number of attribute channels exceeds this (\eg, one atlas for albedo (RGB) and roughness (A), another for tangent normals (RGB) and metallic (A)).

\paragraph{Hardware Accelerated Texture Sampling}

During test-time rendering, we perform texture sampling using hardware-accelerated filtering:\\
\begin{enumerate} 
\item For a given primitive $k$ and local coordinates $(u,v)$, we map to normalized texture coordinates $(s,t) \in [0,1] \times [0,1]$ as described in the main paper (Section 3.3).\\
\item We retrieve the chart coordinates $(chart\_idx\_x, chart\_idx\_y)$ from the indirection buffer.\\
\item We compute the atlas coordinates: 
\begin{equation} 
\begin{split}
s_{\text{atlas}} = \frac{chart\_idx\_x \cdot T + s \cdot T}{N_{\text{charts}_x} \cdot T}, \\ t_{\text{atlas}} = \frac{chart\_idx\_y \cdot T + t \cdot T}{N_{\text{charts}_y} \cdot T}
\end{split}
\end{equation} 
where $N_{\text{charts}_x}$ and $N_{\text{charts}_y}$ are the total number of charts along the width and height of the atlas, respectively.\\
\item We sample the atlas using hardware-accelerated bilinear filtering with CUDA's `tex2D()` intrinsic: \begin{equation} 
\boldsymbol{a}_k(x,y) = \text{BilinearFilter}(\text{atlas}, s_{\text{atlas}}, t_{\text{atlas}}). \end{equation}
\end{enumerate}
This approach fully leverages the GPU's texture units for filtering, which are significantly faster than software-based sampling approaches. It leverages dedicated texture caching and filtering hardware, significantly speeding up attribute fetching compared to manual interpolation from global memory arrays. By using texture atlases, we avoid the need for multiple texture bindings during rendering, improving performance and reducing API overhead.

\begin{figure}[h!]
    \centering
    \includegraphics[width=\columnwidth]{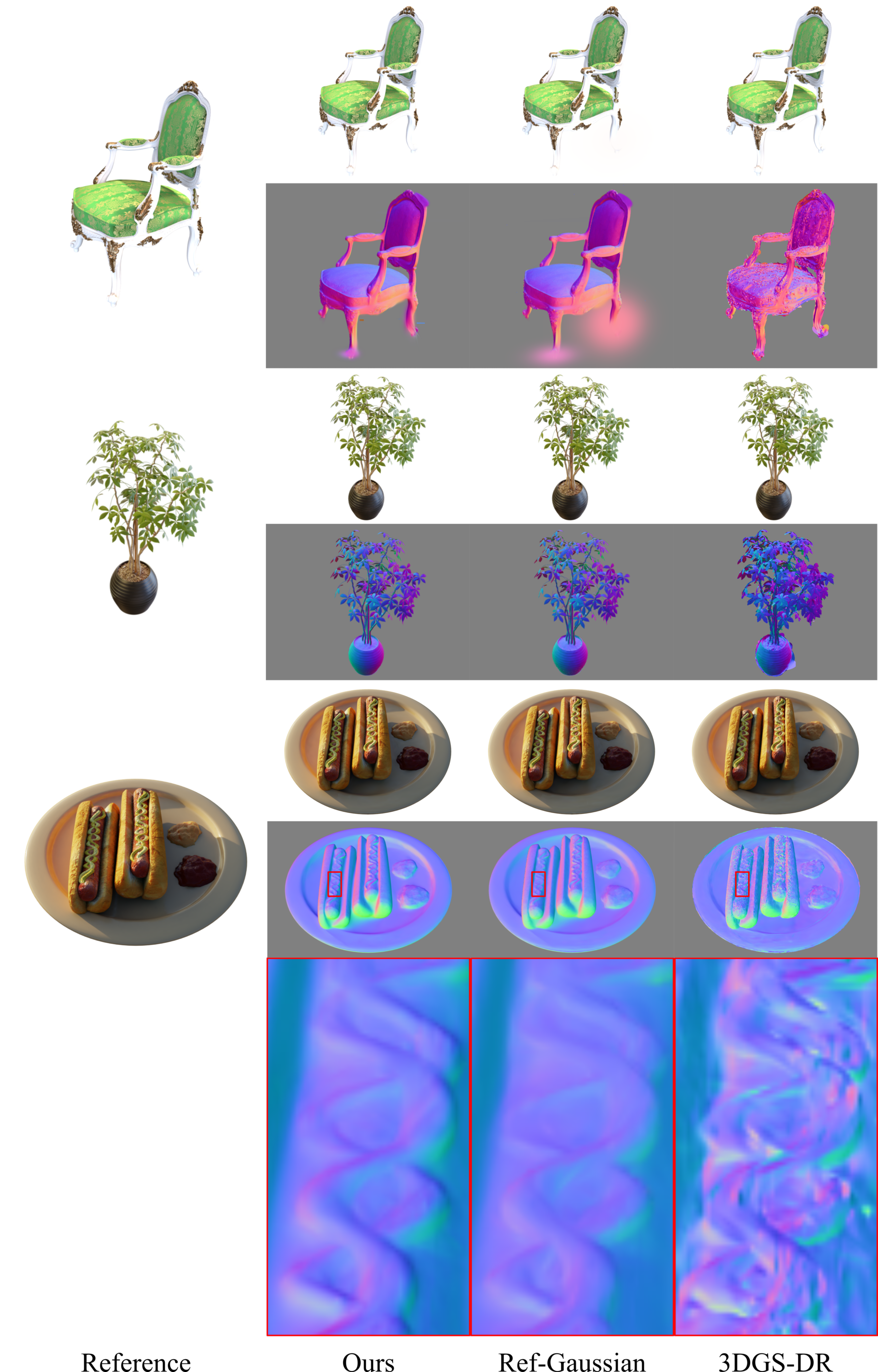}
    \caption{Qualitative comparisons on the Synthetic NeRF dataset. From top to bottom: drums, ficus and hotdog.}
    \label{fig:synth_nerf}
\end{figure}

\begin{table}
\caption{Per-scene image quality comparison for novel view synthesis on NeRF Synthetic dataset.}
\vspace{5pt}
\tabcolsep=0.15cm
\renewcommand\arraystretch{1.3}
\begin{center}
\scalebox{0.8}{
\begin{tabular}{cccccccc}
\hline
              & chair               & drums               & ficus               & hotdog              & lego                & materials           & ship                \\ \hline
\multicolumn{8}{c}{\textbf{PSNR} $\uparrow$}                                                                                                                            \\ \hline
3DGS          & \cellcolor{c1}35.03 & 26.04               & 35.29               & \cellcolor{c2}37.57 & \cellcolor{c1}33.71 & 30.04               & \cellcolor{c1}31.43 \\
3DGS-DR       & 32.10               & 25.31               & 28.03               & 35.58               & 32.94               & 28.35               & 29.07               \\ \hline
Ref-Gaussian   & 33.73               & \cellcolor{c2}26.25 & \cellcolor{c2}35.50 & 37.19               & 32.58               & \cellcolor{c2}30.72 & 30.44               \\
\textbf{Ours} & \cellcolor{c2}34.40 & \cellcolor{c1}26.4 & \cellcolor{c1}35.9 & \cellcolor{c2}37.43 & \cellcolor{c2}33.34 & \cellcolor{c1}31.25 & \cellcolor{c2}30.55 \\ \hline
\multicolumn{8}{c}{\textbf{SSIM} $\uparrow$}                                                                                                                            \\ \hline
3DGS          & \cellcolor{c1}0.987 & \cellcolor{c1}0.954 & \cellcolor{c2}0.987 & \cellcolor{c1}0.985 & \cellcolor{c2}0.975 & 0.959               & \cellcolor{c1}0.906 \\
3DGS-DR       & 0.977               & 0.946               & 0.963               & 0.982               & \cellcolor{c1}0.978 & 0.950               & 0.894               \\ \hline
Ref-Gaussian  & 0.979               & 0.952               & \cellcolor{c2}0.987 & 0.982               & 0.972               & \cellcolor{c2}0.966 & 0.895               \\
\textbf{Ours} & \cellcolor{c2}0.981 & \cellcolor{c2}0.953 & \cellcolor{c1}0.988 & \cellcolor{c2}0.983 & 0.974               & \cellcolor{c1}0.969 & \cellcolor{c2}0.896 \\ \hline
\multicolumn{8}{c}{\textbf{LPIPS} $\downarrow$}                                                                                                                         \\ \hline
3DGS          & \cellcolor{c1}0.012 & \cellcolor{c1}0.040 & \cellcolor{c2}0.012 & \cellcolor{c1}0.021 & \cellcolor{c2}0.027 & 0.040               & \cellcolor{c1}0.111 \\
3DGS-DR       & 0.024               & 0.055               & 0.055               & 0.033               & \cellcolor{c1}0.026 & 0.044               & 0.129               \\ \hline
Ref-Gaussian  & 0.023               & 0.043               & 0.013               & 0.029               & 0.031               & \cellcolor{c2}0.036 & 0.127               \\
\textbf{Ours} & \cellcolor{c2}0.022 & \cellcolor{c2}0.042 & \cellcolor{c1}0.012 & \cellcolor{c2}0.028 & \cellcolor{c2}0.029 & \cellcolor{c1}0.033 & \cellcolor{c2}0.126 \\ \hline
\end{tabular}}
\end{center}
\label{tab:nvs_nerf}
\end{table}

\section{Addtional Results}

\paragraph{Performance on non-specular scenes}
In Table~\ref{tab:nvs_nerf}, we show the performance of our method against different methods in non-specular scenes from the NeRF Synthetic dataset.
Our method is capable of enhancing the baseline method thanks to the expressivity of material and normal texture splatting. It can even reach the level of performance of the vanilla 3DGS~\cite{3dgs} on these scenes.
We also show in Figure~\ref{fig:synth_nerf} comparison against other reflection gaussian splatting methods where the normals produced by our method are more defined and are less noisy in comparison.

\begin{figure}[h!]
    \centering
    \includegraphics[width=\columnwidth]{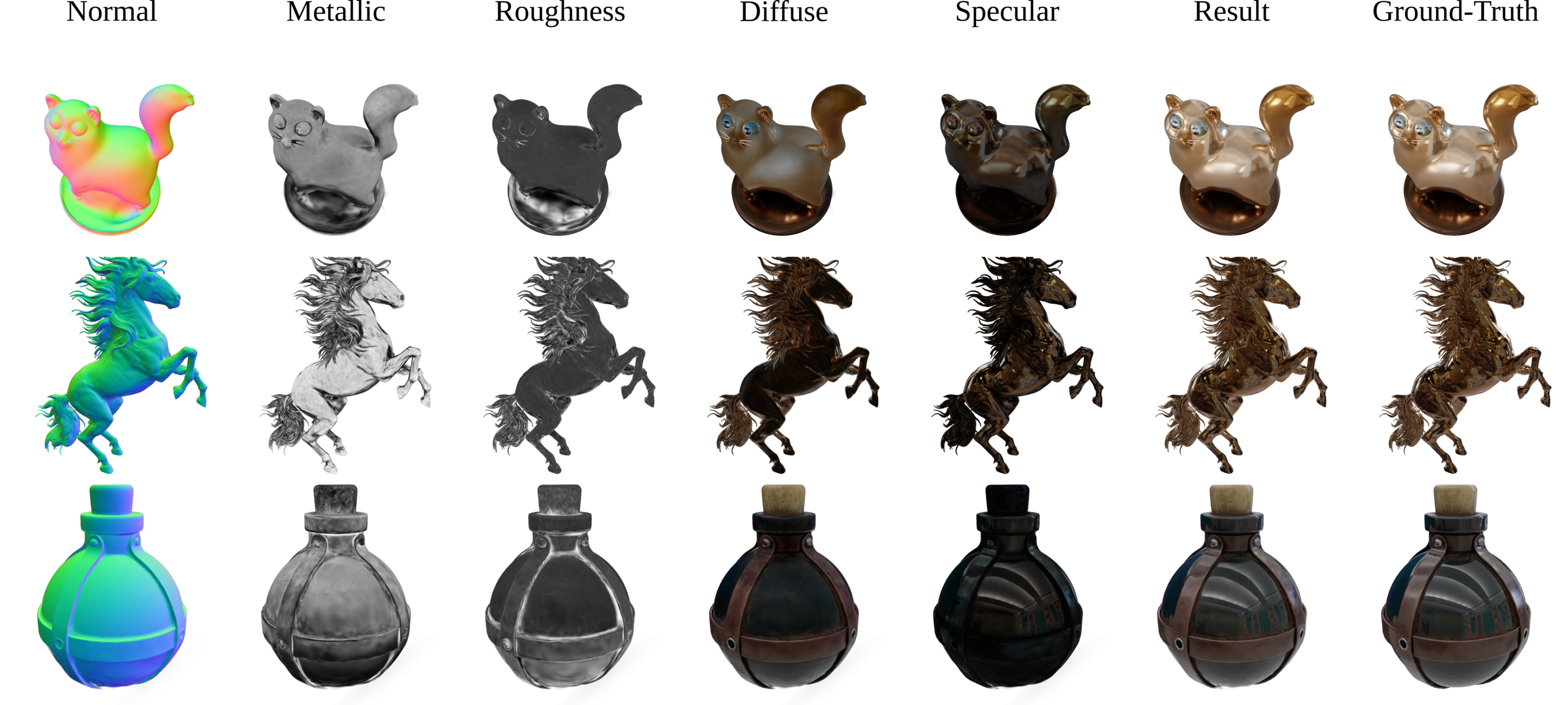}
    \caption{Decomposition results of our method. From top to bottom: cat, horse and potion.}
    \label{fig:decomp}
\end{figure}

\begin{figure}[h!]
    \centering
    \includegraphics[width=\columnwidth]{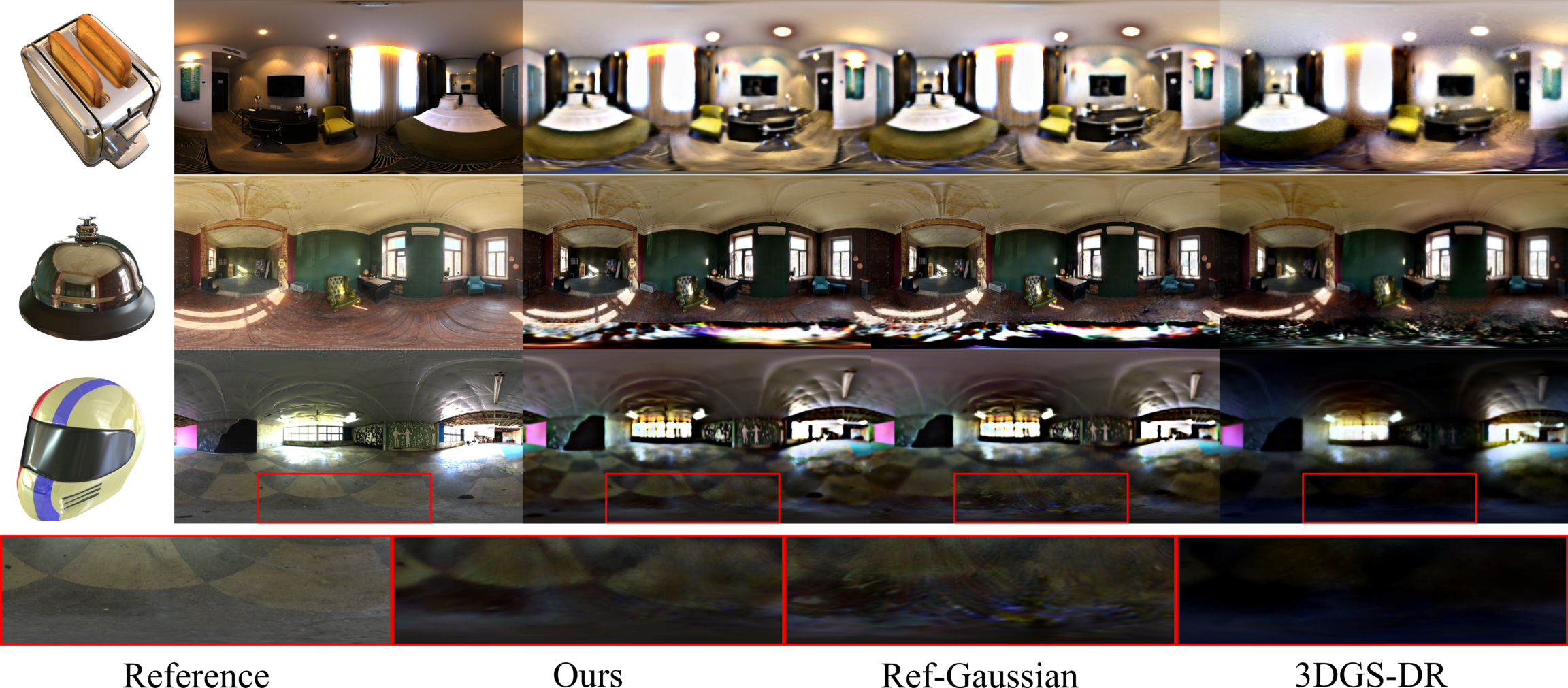}
    \caption{Qualitative comparisons of environment maps estimated by different methods.}
    \label{fig:env}
\end{figure}

\paragraph{Scene Decomposition and Environment Maps}
Despite the under-constrained nature of the task in many cases, our optimization yields stable and physically plausible scene property factorization. Figure \ref{fig:decomp} illustrates decomposition examples for reflective objects into geometry normals, roughness and metallic. We also show the diffuse and specular components, in addition to the final rendering which preserves the ground truth. The specular highlights are mostly successfully factored out from the view-independent component. 

We also compare estimated environment maps for our method and the competition in Figure~\ref{fig:env}. Notice that our estimated scene lighting bares more similarity with the reference maps and displays fewer distortions and artifacts generally, compared to our baseline.

\section{Additional Ablation Studies}

\paragraph{Texture Mapping vs. Increased Primitive Count}
A natural question when evaluating our texture-based method is whether similar quality improvements could be achieved by simply increasing the number of primitives in the baseline method. To investigate this, we compared our approach against baseline models with increased primitive counts.
We created versions of the baseline with 2× and 4× the original number of primitives by lowering the gradient threshold for densification and increasing the frequency of cloning and splitting operations during optimization. We compare against our method that  maintain the original primitive count with the 2×2 textures per primitive. Notably, even with these textures, our method has lower storage requirements than the 2× primitive count baseline as it only incurs an increase of 20\% in storage compared to the baseline method. 
As shown in Table~\ref{tab:primitive_count}, the performance of the baseline method actually degrades as the number of primitives increases beyond the optimal point. This seemingly counter-intuitive result can be explained by the fact that in practice, increasing the number of primitives creates a more complex optimization and that more primitives require more extensive sorting operations during rendering, which can lead to primitives being placed in suboptimal orders, especially with the use of global sorting~\cite{3dgs}. Primitives can also become cluttered, particularly in areas with complex geometry, leading to degraded normal estimation quality which is critical for reflective scenes.

Our texture-based approach avoids these issues by increasing representational power without increasing the number of primitives, leading to better rendering quality particularly for reflective surfaces.

\begin{table}
\caption{Ablation study: Effect of increasing primitive count for the baseline method on average image quality. We evaluate on the Shiny Blender dataset~\cite{ref_nerf}.}
\vspace{10pt}
\tabcolsep=0.15cm
\renewcommand\arraystretch{1.2}
\begin{center}
\resizebox{1.0\columnwidth}{!}{
\begin{tabular}{c|cccc}
\hline
Metric & Ours & Ref-Gaussian & $\times$2 Primitives & $\times$4 Primitives \\ \hline
PSNR $\uparrow$ & \cellcolor{c1}35.90 & \cellcolor{c2}34.89 & 33.63 & 34.12 \\
SSIM $\uparrow$ & \cellcolor{c1}0.977 & \cellcolor{c2}0.974 & 0.967 & 0.970 \\
LPIPS $\downarrow$ & \cellcolor{c1}0.047 & \cellcolor{c2}0.053 & 0.061 & 0.057 \\
\hline
\end{tabular}
}
\label{tab:primitive_count}
\end{center}
\end{table}

\begin{figure}[h!]
    \centering
    \includegraphics[width=\columnwidth]{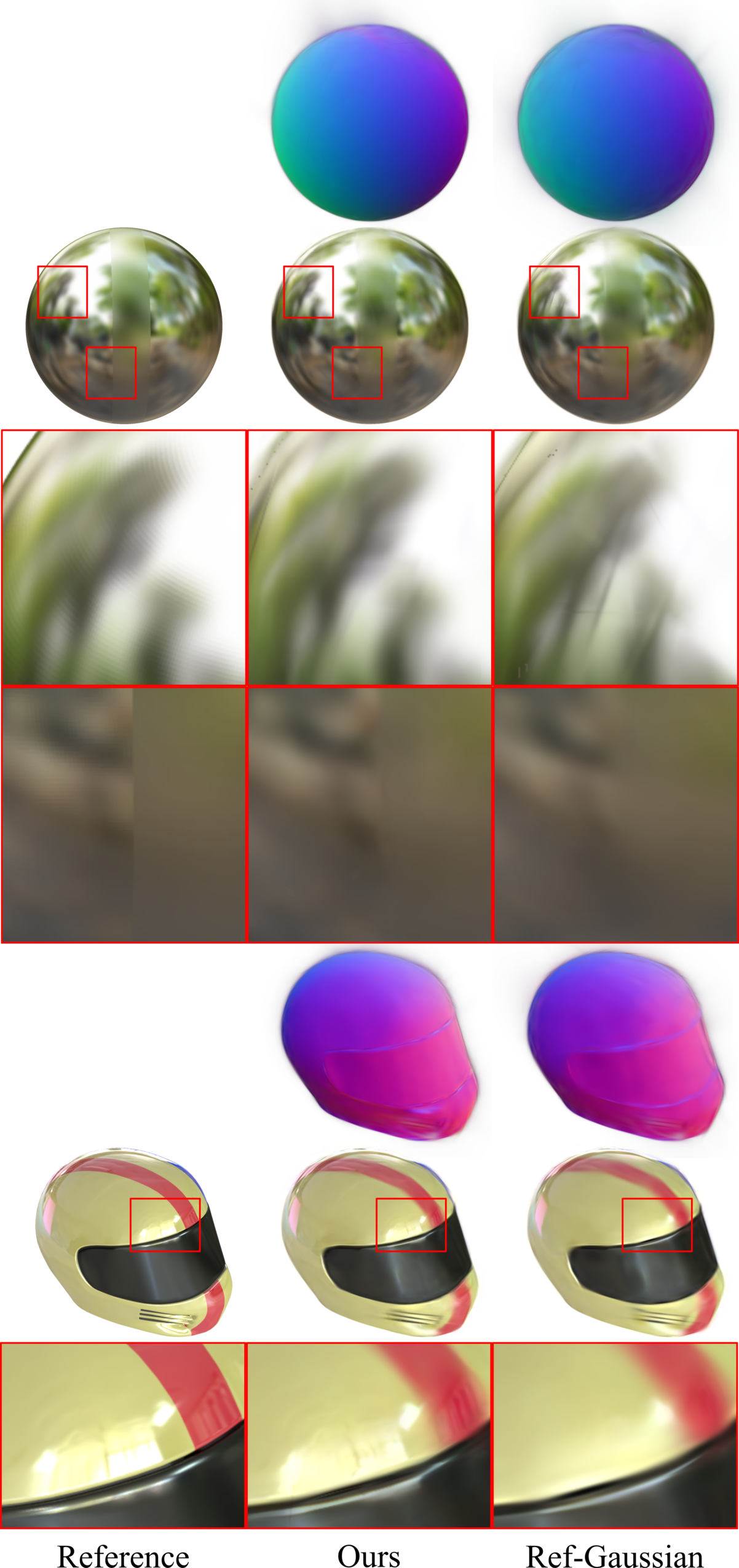}
    \caption{Comparison of novel view synthesis and normal rendering with fewer primitives}
    \label{fig:reduced_primitives}
\end{figure}

\paragraph{Performance with Reduced Primitive Count}
To further demonstrate the efficiency of our texture-based representation, we conducted experiments with reduced primitive counts. We created models using only 50\% of the primitives used in our standard optimization process, for both the baseline method and our texture-based approach.

As illustrated in Figure~\ref{fig:reduced_primitives}, the baseline method with reduced primitive count suffers from significant degradation in normal reconstruction quality, exhibiting holes and artifacts due to the limited capacity of the model. In contrast, our texture-based method maintains smoother normal reconstruction even with the reduced primitive count, demonstrating the expressiveness of per-primitive texture mapping.

\end{document}